\input harvmac

\Title{ \vbox{\baselineskip12pt
\hbox{hep-th/9906070}
\hbox{HUTP-99/A034}
\hbox{IASSNS-HEP-99-52}
\hbox{PUPT-1864}}}
{{\vbox{\centerline{CFT's From Calabi-Yau Four-folds }}}}
\smallskip
\centerline{Sergei Gukov}
\smallskip
\centerline{\it Department of Physics, Princeton University}
\centerline{\it Princeton, NJ 08544, USA}
\smallskip
\centerline{Cumrun Vafa}
\smallskip
\centerline{\it Jefferson Laboratory of Physics}
\centerline{\it Harvard University, Cambridge, MA 02138, USA}
\centerline{Edward Witten}
\smallskip
\centerline{\it School of Natural Sciences, Institute for Advanced Study}
\centerline{\it Olden Lane, Princeton, NJ 08540, USA}\bigskip

\bigskip

\def\tilde{\widetilde}
\def\hat{\widehat}
\def\bar{\overline}
\vskip .3in
We consider F/M/Type IIA theory compactified to four, three,
or two dimensions
on a Calabi-Yau four-fold, and study the behavior near an isolated
singularity   in the presence of appropriate fluxes and branes.
We analyze the vacuum and soliton structure of these models, and show
that near an isolated singularity, one often
 generates massless chiral superfields and a superpotential,
and in many instances in two or three dimensions
one obtains nontrivial superconformal field theories.
  In the case of two dimensions,
we identify some of these theories with certain Kazama-Suzuki coset models,
such as the $ {\cal N}=2$ minimal models.

\def\underarrow#1{\vbox{\ialign{##\crcr$\hfil\displaystyle
 {#1}\hfil$\crcr\noalign{\kern1pt\nointerlineskip}$\longrightarrow$\crcr}}}
\Date{June 1999}

\newsec{Introduction}

We have learned in recent years that it is fruitful to study
 singular limits of string compactifications.  In this
paper, we consider  theories with four supercharges
in four, three, and two
 dimensions, constructed by considering $F$-theory, $M$-theory, and
Type IIA string theory
on a Calabi-Yau four-fold with an isolated complex singularity.
We can connect these theories to each other by circle compactifications
from four to three to two
 dimensions.  In addition to the choice of singularity,
the description of these theories depends on certain additional
data involving the four-form flux and membrane charge
in $M$-theory (and related objects in the other theories).

We will analyze the
vacuum structure of these theories
and the domain walls connecting the possible vacua.
We argue that in many cases,
the nonperturbative physics near a singularity generates massless
chiral superfields with a superpotential, leading in many instances,
especially in two dimensions or in three dimensions with large membrane
charge,
to an infrared flow to a nontrivial conformal field theory.
In some cases, we can identify the theories in question with known
superconformal models; for example, Type IIA at a four-fold $A_n$ singularity
gives an ${\cal N}=2$ Kazama-Suzuki model in two dimensions, as we argue using
the
vacuum and soliton
structure.
More generally, from the A-D-E four-fold hypersurface singularities
with appropriate fluxes, we obtain all the ${\cal N}=2$ Kazama-Suzuki models
\ref\ksu{Y. Kazama and H. Suzuki,``New N=2 Superconformal Field
Theories and Superstring Compactification,''
Mod. Phys. Lett. {\bf A4} (1989) 325.}\ at level one.  This is a large
list of exactly solvable conformal theories, which includes
the ${\cal N}=2$ unitary minimal models. It is quite satisfying
that strings in the presence of singularities captures such
a large class of known
conformal theories
in two dimensions, and suggests that
maybe even in higher dimensions, strings propagating in
singular geometries yield an
equally
large subspace of conformal theories.  Moreover,
since the Kazama-Suzuki
models are exactly solvable conformal theories in two dimensions,
it would be interesting to see to what extent its known
spectrum and correlation
functions can be extracted
from string theory.  This would be a natural
testing grounds in view of potential application to
higher dimensions where
the conformal theories are less well understood.

Our result gives a relation between
singularity theory, as in the Landau-Ginzburg
approach to conformal theories in two dimensions \nref\zam{
A.B. Zamolodchikov, JETP Lett. {\bf 46} (1987) 160.}\nref\vwa{
C. Vafa and N.P. Warner,``Catastrphes and the Classification of
Conformal Theories,'' Phys. Lett. {\bf B218} (1989) 51.
}\nref\mart{E. Martinec,``Algebraic Geometry and Effective
Lagrangians,'' Phys. Lett. {\bf B217} (1989) 431.}\
\refs{\zam - \mart}, and the singularity
of internal compactification geometry.
This relation may well extend to non-supersymmetric examples;
it would certainly be interesting to explore this.

In  section 2,  we analyze the fluxes, branes, and vacuum
states near a four-fold singularity.  In section 3, we show how to
compute the spectrum of domain walls (which can also be viewed
as strings and kinks in the three and two-dimensional cases) for a
special class of singularities.    In section
4,  we identify the models derived from A-D-E singularities
with Kazama-Suzuki models at level 1.  In section 5, we discuss
some additional classes of singularities on four-folds, and in
section 6, we discuss the reinterpretation of some of our results
in terms of branes.

\newsec{Classification Of Vacua}

\subsec{The $G$-Field And Domain Walls}
\def\R{{\bf R}}
\def\Z{{\bf Z}}

For our starting point, we take $M$-theory on $\R^3\times Y$,
where $Y$ is a compact eight-manifold.  Soon, we will specialize
to the case that $Y$ is a Calabi-Yau four-fold, so as to achieve supersymmetry.
We also will note in section 2.5 the generalization of our remarks to
Type IIA or $F$-theory compactification on $Y$.

To fully specify a vacuum on $Y$, one must specify not just $Y$ but
 also the topological
class of the three-form potential $C$ of $M$-theory, whose field
strength is $G=dC$.  Roughly speaking, $C$-fields are classified
topologically by a characteristic class  $\xi\in H^4(Y;\Z)$.
At the level of de Rham cohomology, $\xi$ is measured by the differential
form $G/2\pi$, and we sometimes write it informally as $\xi=[G/2\pi]$.
\foot{The assertion that $\xi$ takes values in $H^4(Y;\Z)$
is a bit imprecise, since in general
\ref\ewitten{E. Witten, ``On Flux Quantization In $M$-Theory And The Effective
Action,'' J. Geom. Phys. {\bf 22} (1997) 1.}\ the $G$-field is shifted from
standard Dirac quantization
and $\xi$ is not an element of
$H^4(Y;\Z)$.  But the difference between two $C$-fields is always measured
by a difference $\xi-\xi'\in H^4(Y;\Z)$.  $\xi$ itself takes values in
a ``principal homogeneous space'' $\Lambda$ for the group $H^4(Y;\Z)$;
 the relation between $H^4(Y;\Z) $ and $\Lambda$ is just
analogous to the relation between $H^2(Y;\Z)$ and the set of ${\rm Spin}_c$
structures on $Y$.  The shift in the quantization law of $G$ arises
precisely when the intersection form on $H^4(Y;\Z)$ is not even.
In our examples, this will occur only in section 5, and we will ignore
this issue until that point.}

Without breaking the three-dimensional Poincar\'e symmetries, this
model can be generalized by picking $N$ points $P_i\in Y$
and including $N$ membranes  with world-volumes
of the form $\R^3\times P_i$.
More generally, we include both membranes and antimembranes and
let $N$ be the difference between the number of membranes and antimembranes;
thus it can be a positive or negative integer.
With $Y$ being compact,  the net source of the $C$-field must vanish;
this gives
a relation \nref\sethietal{S. Sethi, C. Vafa, and E. Witten,
``Constraints on Low Dimensional String Compactifications,''
Nucl. Phys. {\bf B480} (1996) 213.}%
\refs{\sethietal,\ewitten}
\eqn\sethrel{N={\chi\over 24}-{1\over 2}\int_Y {G \wedge G\over (2\pi)^2}.}
If $G$ obeys the shifted quantization condition mentioned in the last footnote,
then the right hand side of \sethrel\ is always integral \ewitten.

In this construction, models defined using the
same $Y$ but different $\xi$ are actually different states of the same
model.\foot{This may also be true of models with
 different $Y$ -- as suggested by results of
\ref\morr{B. Greene, D. Morrison, and A. Strominger,``Black
hole Condensation and the Unification of String Vacua,'' Nucl.
Phys. {\bf B451} (1995) 109.} in the
threefold case -- but this issue is much harder to explore.}
To show this, it suffices to describe a domain wall interpolating
between models with the same $Y$ and with $C$-fields of arbitrary
characteristic classes $\xi_1$ and $\xi_2$.
By Poincar\'e duality, $H_4(Y;\Z)=H^4(Y;\Z)$.
Hence, there is a four-cycle $S\subset Y$, representing an element of
$H_4(Y;\Z)$,
such that if $[S]\in H^4(Y,\Z)$ is the class that is Poincar\'e dual
to $S$, then $\xi_2-\xi_1=[S]$.  Now, consider a fivebrane in $\R^3\times
Y$ whose worldvolume is $W=\R^2\times S$, with $\R^2$ a linear subspace
of $\R^3$.  Being of codimension one in spacetime, this fivebrane looks
macroscopically like a domain wall.  Moreover, because the fivebrane
is a magnetic source of $G$, the characteristic class $\xi=[G/2\pi]$
jumps by $[S]$ in crossing this domain wall.  Hence if it equals $\xi_1$
on one side, then it equals $\xi_2=\xi_1+[S]$ on the other side.

Equation \sethrel\ implies that if $\xi$ jumps in crossing a domain
wall, then $N$ must also jump.  Let us see how this comes about.
The key is that there is a self-dual three-form $T$ on the fivebrane
with a relation
\eqn\kixo{dT= G\vert_W -2\pi\delta(\partial M).}
where $G\vert_W$ is the restriction of the $G$-field to the world-volume $W$,
$\partial M$ is the union of all boundaries of membrane worldvolumes
that terminate on $W$, and $\delta(\partial M)$ is a four-form
with delta function support on $\partial M$.
Because the $G$-field actually jumps in crossing
the fivebrane, it is not completely obvious how to interpret the term $G|_W$.
We will assume that this should be understood as the average of the
$G$-field on the two sides:
 $G|_W=(G_1+G_2)/2$.  Since the left hand side of \kixo\ is
zero in cohomology, we get a relation in cohomology
\eqn\pixo{[\partial M] ={G_1+G_2\over 2(2\pi)}}
where $[\partial M]$ is the cohomology class dual to $\partial M$.
We are interested in the case that the membrane worldvolumes are of the
form $\R^3\times P_i$, so that their boundaries on $W$ are of the form
$\R^2\times P_i$.  In evaluating \pixo, we can suppress the common
$\R^2$ factor, and integrate over $S$ to get a cohomology relation.
The integration converts $[\partial M]$ into $N_1-N_2$, the change
in $N$ in crossing the domain wall or in other words the net number
of membranes whose boundary is on the fivebrane.
So
\eqn\ujixo{N_1-N_2={1\over 2}\int_S{G_1+G_2\over 2\pi}=-{1\over 2}\int_Y
{G_1^2-G_2^2\over (2\pi)^2}.}
Here we have used the fact that $[S]=[(G_2-G_1)/2\pi]$ to convert an
integral over $S$ to one over $Y$.
Clearly, \ujixo\ is compatible with the requirement that \sethrel\
should hold on both sides of the domain wall.

\bigskip\noindent{\it Incorporation Of Supersymmetry}

Now we wish to specialize to the case that $Y$ is a Calabi-Yau
four-fold and to look for vacua with unbroken supersymmetry.

For this, several restrictions must be imposed.  The requirements
for $G$, assuming that one wants unbroken supersymmetry with zero
cosmological constant, have been obtained in an elegant
computation \ref\beckers{K. Becker and M. Becker,``M Theory on Eight
Manifolds,'' Nucl. Phys. {\bf B477} (1996) 155.}.
  The result is that $G$ must be of type $(2,2)$ and
must be primitive, that is, it must obey
\eqn\tucico{K\wedge G=0,}
where $K$ is the K\"ahler form.  We will analyze this condition in  Appendix I,
but for now we note that it implies that $G$ is self-dual and
hence in particular that
\eqn\olivo{\int_Y{G\wedge G\over (2\pi)^2} \geq 0,}
with equality only if $G=0$.

The second basic consequence of supersymmetry is that $N$ must
be positive.
Only membranes and not anti-membranes on $\R^3\times P_i$
preserve the same supersymmetry
that is preserved by the compactification on the complex four-fold $Y$.

Given that $N$ must be positive, the relation \sethrel\ implies that
\eqn\ixo{\int_Y{G\wedge G\over (2\pi)^2}\leq {\chi\over 12}.}
This inequality together with self-duality implies that,
for compact $Y$, there are only finitely many choices
of $G$ that are compatible with unbroken supersymmetry.  For $\chi$
negative, there are none at all.

\bigskip\noindent{\it Energetic Considerations}

There is another way to understand the result that
$\Phi=N+ {1\over 8\pi^2}\int G\wedge G$ should
not change in crossing the domain wall.  This is based on a finite energy
condition. The condition that the domain wall be flat and
have finite tension
requires that the energy density on the two sides of the domain wall
should be equal far away from the domain wall.  The energy density
in the bulk gets contribution from the $G$ flux given by
${1\over 8 \pi ^2}\int G\wedge *G$ and from the membranes
by $N$.  For the supersymmetric situation we are considering,
$G$  is self-dual (as explained in  Appendix I),
i.e., $G=*G$ so the energy density is given by
$N+{1\over 8 \pi ^2}\int G\wedge G$, and so its constancy
across a domain wall is a consequence of the finite energy of the
domain wall.  This is also important in our applications later,
as we will use a BPS formula for the mass of domain walls.  Due to
boundary terms at infinity, such
formulas are generally not valid for objects of very low codimension
in space (the codimension is one in our case).  The cancellation of the
boundary terms in question in  our case is again precisely
the condition of constancy of $\Phi$ across the domain wall.

\subsec{Behavior Near  A Singularity}

So far we have considered the case of a compact smooth manifold $Y$.
Our main interest in the present paper, however, is to study the
behavior as $Y$ develops a singularity.  For practical purposes,
it is convenient then to omit the part of $Y$ that is far from the
singularity and to consider a complete but not compact Calabi-Yau
four-fold that is developing a singularity.
In fact, some of the singularities we will study -- like the
$A_n$ singularities of a complex surface for very large $n$ -- probably
cannot be embedded in a compact Calabi-Yau manifold.
Our discussion will apply directly to an isolated singularity of a non-compact
variety.

Hypersurface singularities are an important example and will be our
focus in this paper except in section 5.  For example, one of our
important applications will be to quasihomogeneous hypersurface singularities.
In this case, we begin with five complex variables $z_a$, $a=1,\dots, 5$
of degree $r_a>0$ and a polynomial $F(z_1,\dots,z_5)$ that is homogeneous
of degree 1.   We assume that $F$ is such that the hypersurface $F=0$
is smooth except for an isolated singularity at $z_1=\dots =z_5=0$.
Then we let $X$ be
a smooth deformation of this singular hypersurface such as
\eqn\turfox{F(z_1,\dots,z_5)=\epsilon}
with $\epsilon$ a constant, or more generally
\eqn\burfox{F(z_1,\dots,z_5)=\sum_it_iA_i(z_1,\dots,z_5),}
with complex parameters $t_i$ and polynomials $A_i$ that describe
relevant perturbations of the singularity $F=0$.
The singular hypersurface $F=0$ admits the $U(1)$ symmetry
group
\eqn\junon{z_a\to e^{i\theta r_a}z_a.}
Under this transformation, the holomorphic four-form
\eqn\omegga{\Omega={dz_2\wedge dz_3\wedge dz_4\wedge dz_5\over
{\partial F/\partial z_1}}}
has charge
\eqn\rommega{r_\Omega=\sum_a r_a-1.}
The $U(1)$ symmetry in \junon\
is an $R$-symmetry group if $r_\Omega\not= 0$.
If the model is to flow in the infrared to a superconformal field theory,
an $R$-symmetry must appear in the superconformal algebra; we propose
that it is the symmetry just identified.  If the $A_i$ have degrees
$r_i$, then the dimensions of the corresponding operators are
proportional to $r_i/r_\Omega$ (in other words, the $R$-charges normalized
so that $\Omega$ has $R$-charge 1).  Since the $r_i$ are positive,
requiring that the dimensions be positive gives a condition
$r_\Omega>0$:
\eqn\ilbo{\sum_ar_a>1.}
We will see the importance of this condition from several points of
view.

In what sense is such an $X$ a Calabi-Yau manifold?
\nref\tyau{G. Tian and S.-T. Yau, ``Complete K\"ahler Manifolds With
Zero Ricci Curvature II, Inv. Math. {\bf 106} (1991) 27.}%
The holomorphic four-form $\Omega$ defined in \omegga\
has no zeroes or poles on $X$, though it has in a certain sense
(using the compactification described in the next paragraph) a pole at
infinity.
 A theorem
of Tian and Yau \tyau\ asserts, assuming \ilbo,
that there is a Calabi-Yau metric on $X$
with volume form
\eqn\ycoo{|\Omega\wedge \bar\Omega|,}
and moreover (see the precise statement in eqn. (2.3) of \tyau) this
metric is asymptotically conical, that is it looks near infinity
like \eqn\sni{ds^2=dr^2+r^2ds_\perp^2.}
Here $ds_\perp$ is an ``angular'' metric, and $r$ is a ``radial''
coordinate near infinity which scales   under $z_a\to \lambda^{r_a}z_a$
as $r\to \lambda^{\left(\Sigma_ar_a-1\right)/4}r$.  This exponent
  ensures that the
volume form derived from \sni\ scales like $|\Omega|^2$.

To apply the Tian-Yau theorem to the hypersurface $X$ and deduce the
statements in the last paragraph,
one writes $r_a=b_a/d$ with $b_a$ relatively prime integers and $d$
a positive integer.  Then one introduces another complex variable
$w$ of
degree $1/d$, and one compactifies $X$ to the compact variety
$Y'$ defined by the equation
$F(z_i)-\epsilon w^d=0$ in a weighted projective space.
The discussion of \tyau\ applies to this
situation, with $D$ the divisor $w=0$, and identifies $r$ with
a fractional power of $|w|$.

It is very plausible that if a compact Calabi-Yau manifold $Y$
develops an isolated hypersurface singularity that is at finite
distance on the moduli space, then the Calabi-Yau metric on $Y$
looks locally like the conelike metric that we have just described on
the hypersurface $X$.\foot{There is no claim here that $X$ can be globally
embedded in $Y$, only that the behavior of $Y$ near its singularity
can be modeled by $X$.  Note that the variety $Y'$ used in the last
paragraph in relation to the Tian-Yau theorem is not a Calabi-Yau manifold.}
For our purposes, we do not strictly need to know that this is true,
but the physical applications are certainly rather natural if it is.

\bigskip\noindent{\it Flux At Infinity}

Noncompactness of $X$ leads to several important novelties in the
specification of the model.
First of all, flux can escape to infinity, and hence \sethrel\ no longer
holds.  Rather, an extra term appears in \sethrel, namely the flux
$\Phi$ measured at infinity.  This flux is a constant of the motion,
invariant under the dynamics which occurs in the ``interior'' of $X$.
If we absorb the constant $\chi/24$ in the definition of $\Phi$,\foot{When
$X$ is not compact, $\chi$ must in any event be defined by
a curvature integral and need not coincide with
 the topological Euler characteristic.}
then we can write the conserved quantity as
\eqn\nuffy{\Phi=N+{1\over 2}\int_X{G\wedge G\over (2\pi)^2}.}
We can think of $\Phi$ as a constant that must be specified (in addition
to giving $X$) in order to determine the model.  A model with given
$\Phi$ has various vacuum states, determined by the values of $N$
and $G$.  For unbroken supersymmetry, both terms on the right hand
side of  \nuffy\ are positive (for the same reasons as in the case
of compact $X$), so there are only finitely many possible choices
of $N$ and $G$ for fixed $\Phi$.

In addition to $\Phi$, there is another quantity that characterizes
the definition of the model -- and commutes with the dynamics.
For finiteness of the energy, it is reasonable to require that
the flux $G$ vanishes if restricted to $\partial X$, the region
near infinity in $X$.  This does not imply that the cohomology class
$\xi$ vanishes if restricted to $\partial X$, but only that its
restriction is a torsion class.  Thus,  the $C$-field
at infinity is flat, but perhaps topologically non-trivial.
Local dynamics cannot change the behavior at infinity, so the
restriction of $\xi$ to $\partial X$ is another invariant of the local
dynamics, which must be specified in defining a model.

There is another way to see more explicitly how this invariant comes about.
For this, we have to look at precisely what Poincar\'e duality says
in the case of a noncompact manifold $X$.  Domain walls of the
type introduced in section 2.1 are classified
by $H_4(X;\Z)$, which classifies the cycles $S$ on which a fivebrane
can wrap to make a domain wall.
Poincar\'e duality says that this is the same as $H^4_{cpct}(X;\Z)$
(where $H^4_{cpct}$ denotes cohomology with compact support),
\foot{The cohomology
of $X$ with compact support is generated by closed forms $\beta$ on
$X$ with compact support, subject to the equivalence relation
that $\beta\cong \beta+d\epsilon$ if $\epsilon$ has compact support.}
the rough
idea being that if $S$ is a four-cycle determining an element of
$H_4(X;\Z)$ (so in particular $S$ is by definition compact),
then the Poincar\'e dual class $[S]$ is represented
by a delta function $\delta(S)$ that has compact support.
$C$-fields on $X$ are classified topologically by $\xi \in H^4(X;\Z)$.
The groups $H^4(X;\Z)$ and $H^4_{cpct}(X;\Z)$ that classify, respectively,
topological classes of $C$-fields and of changes in $C$-fields in
crossing domain walls are in general different for non-compact $X$.
However,
there  is always
a natural map
\eqn\koop{i:H^4_{cpct}(X;\Z)\to H^4(X;\Z)}
 (by ``forgetting''
that a class has compact support).
Moreover, for hypersurface singularities,
$H^4_{cpct}(X;\Z)$ and $H^4(X;\Z)$ are lattices,
which we will call $\Gamma$ and $\Gamma^*$ respectively. Poincar\'e duality
in the noncompact case says that $\Gamma$ and $\Gamma^*$ are dual
lattices.
When the intersection form on $H^4_{cpct}(X;\Z)$ has no null vectors,
the map $i$ is an embedding, and $\Gamma$ can be regarded as a sublattice
of its dual lattice $\Gamma^*$.
This makes things very simple.

The lattice $\Gamma$ can actually be described rather simply.
In fact, topologically,
the hypersurface $X$ is homotopic to a ``bouquet'' of four-spheres.
\foot{Such a bouquet is, by definition, associated with
 a tree diagram in which the vertices
represent four-spheres and two vertices are connected by a line if and
only if the corresponding four-spheres intersect.  Such a diagram
has the form of a simply-laced Dynkin diagram (with vertices for
four-spheres and lines for intersections of them), except
that the Cartan matrix may not be positive definite and thus one is
not restricted to the A-D-E case.}
 $H_4(X;\Z)$ is a lattice $\Gamma$
with one generator for every four-sphere in the bouquet.

In crossing a domain wall,
$\xi$ cannot change by an arbitrary amount, but only by something of the
form $i([S])$ where $[S]$ is a class with compact support.
The possible values of $\xi$ modulo changes due to the dynamics, that
is due to crossing domain walls, are thus classified by
\eqn\kilxp{H^4(X;\Z)/i(H^4_{cpct}(X;\Z))=\Gamma^*/\Gamma.}

This can be reinterpreted as follows.
The exact sequence of the pair $(X,\partial X)$ reads in part
\eqn\ollo{\dots H^4(X,\partial X;\Z)\underarrow{i} H^4(X;\Z)\underarrow{j}
H^4(\partial X;\Z)\to H^5(X,\partial X;\Z)\to \dots.}
Here $H^i(X,\partial X;\Z)$ is the same as $H^i_{cpct}(X;\Z)$.
By Poincar\'e duality, $H^5(X,\partial X;\Z)=H_3(X;\Z)$, and this is
zero on dimensional grounds for a bouquet of four-spheres.  So the
exact sequence implies that
\eqn\otollo{H^4(\partial X;\Z)=H^4(X;\Z)/i(H^4_{cpct}(X;\Z))=\Gamma^*/\Gamma.}
Thus the value of $\xi$ modulo changes in crossing domain walls
(the right hand side of \otollo) can be identified with the restriction
of $\xi$ to $\partial  X$ (the left hand side).

If the intersection pairing on $H_4(X;\Z)$ is degenerate, then we should
define $\Gamma$ to be the quotient of $H^4(X,\partial X;\Z)$ by the
group of null vectors (which can be shown to be precisely the image
in $H^4(X,\partial X;\Z)$ of $H^3(\partial X;\Z)$).  The dual $\Gamma^*$
is the subgroup of $H^4(X;\Z)$ consisting of elements whose restriction
to $\partial X$ is torsion.  With these definitions of $\Gamma$ and
$\Gamma^*$, everything that we have described above carries over
($i$ embeds $\Gamma$ as a finite index sublattice of $\Gamma^*$;
 the $G$-fields, with appropriate boundary
conditions, take values in $\Gamma^*$, and jump in crossing a domain
wall by elements of $\Gamma$).

\bigskip\noindent{\it Examples}

We will now illustrate these perhaps slightly abstract ideas
with examples that will be important later.

Consider first the simple case that $X$ is a deformation of
a quadric singularity:
\eqn\olbo{\sum_{a=1}^5 z_a^2 = \epsilon.}
If we assume that $\epsilon$ is real and write $z_a=x_a+iy_a$,
we get $\vec x^2 - \vec y^2=\epsilon$ and $\vec x\cdot \vec y=0$.
Setting $\vec u = \vec x/\sqrt{\epsilon+\vec y^2}$, we see that
$\vec u$ is a unit vector.  The subset of $X$ with $\vec y=0$ is
a four-sphere $S$; since $\vec y\cdot \vec u=0$,
$X$ is the cotangent bundle of $S$.  In particular,
$X$ is homotopic to the four-sphere $S$.
This is the case in which the bouquet of spheres is made from just
a single sphere.
The self-intersection
number of $S$ is $S\cdot S=2$.\foot{To compute this, deform $S$
to the four-sphere $S'$ defined by $y_1=u_2,$ $y_2=-u_1$, $y_3=u_4, $
$y_4=-u_3$, $y_5=0$.  Then $S'$ intersects $S$ at the two points
$u_1=\dots =u_4=0$, $u_5=\pm 1$, and each point contributes $+1$
to the intersection number.  Hence $S\cdot S=S\cdot S'=2$.}
The lattice $\Gamma=H^4_{cpct}(X;\Z)$ is generated by $[S]$,
but the dual lattice $\Gamma^*=H^4(X;\Z)$ is generated by $\half [S]$
(whose scalar product with $S$ is 1).
So $H^4(\partial X;\Z)=H^4(X;\Z)/H^4_{cpct}(X;\Z)=\half\Gamma/\Gamma=\Z_2$.

A somewhat more sophisticated example is the $A_{n-1}$ singularity
in complex dimension four:
\eqn\nolbo{P_n(z_1) +\sum_{a=2}^5 z_a^2=0.}
Here $P_n(z_1)$ is a polynomial of degree $n$.  For simplicity
we take
\eqn\holgo{P_n(z_1)=\prod_{i=1}^n(z_1-b_i)}
with real $b_i$, $b_1<b_2<\dots <b_n$.
For $i=1,\dots, n-1$, we define a four-sphere $S_{i}$ by requiring
that $z_1$ is real with $b_i<z_1<b_{i+1}$, and that the $z_j$ for
$j>1$ are all real or all imaginary depending on the value of $i$ modulo
two.  The $S_i$ generate the lattice $\Gamma=H^4_{cpct}(X;\Z)$.
The intersection numbers of the $S_i$ are $S_i^2=2$,
$S_i\cdot S_{i+1}=1$, with others vanishing. ($S_i$ intersects $S_{i+1}$
at the single point $z_1=b_{i+1}$, $z_j=0$ for $j>1$; $S_i$ does not intersect
$S_j$ if $|j-i|>1$.) Endowed with this intersection
form, $\Gamma$ is the root lattice of the Lie group $A_{n-1}=SU(n)$,
while the dual lattice $\Gamma^*=H^4(X;\Z)$
is the weight lattice.  The quotient is $H^4(\partial X;\Z)=\Gamma^*/\Gamma
=\Z_n$.  It can be shown that $X$ is homotopic to the union of the $S_i$,
which form a ``bouquet.'' In this case, the bouquet is associated
with the Dynkin diagram of $A_n$.

More generally, if $H(z_1,z_2,z_3)$ is a polynomial in three complex
variables that describes a deformation of an A-D-E surface singularity,
we can consider the corresponding surface singularity in complex
dimension four:
\eqn\yolgo{H(z_1,z_2,z_3)+z_4^2+z_5^2=0.}
The case just considered, with $H(z_1,z_2,z_3)=P_n(z_1)+z_2^2+z_3^2$,
corresponds to $A_{n-1}$.  (The appropriate $H$'s for the other cases
are written at the end of section 2.5.) For any of the A-D-E examples,
$\Gamma$ is the root lattice
of the appropriate simply-connected A-D-E group $G$, $\Gamma^*$ is
the weight lattice of $G$, and the quotient $H^4(\partial X;\Z)=
\Gamma^*/\Gamma$ is isomorphic to the center of $G$.
One approach to proving these assertions is to
show that they are true for the middle-dimensional cohomology
of the surface $H(z_1,z_2,z_3)=0$, and are unaffected by
 ``stabilizing''
the singularity by adding two more variables with
the quadratic terms $z_4^2+z_5^2$.

\subsec{Distance To Singularity And Hodge Structure Of Cohomology}

In the present subsection,
we return to the case of a compact Calabi-Yau four-fold $Y$.
We suppose that, when some complex parameters $t_i$ are varied,
$Y$ develops a singularity that looks like a quasihomogeneous hypersurface
singularity $F(z_1,\dots,z_5)=0$, where the $z_a$ have degrees $r_a>0$
and $F$ is of degree 1.
Upon varying the complex structure of $Y$, the hypersurface is deformed
to a smooth one which looks locally like
\eqn\pxol{F(z_1,\dots, z_5)+\sum_it_i A_i(z_1,\dots,z_5) = 0.}
Here the $t_i$ are complex parameters, and the $A_i$ are perturbations
of the equation.

The first question to examine is whether the singularity at $t_i=0$
can arise at finite distance in Calabi-Yau moduli space.
The K\"ahler form on the parameter space is
\eqn\okil{\omega =dt^id\bar t^j{\partial^2 \over \partial t^i\partial \bar t^j}
 K,}
 where $K$ is the K\"ahler potential $K$.  On the parameter space
 of a compact Calabi-Yau manifold, the K\"ahler potential of
the Weil-Peterson metric is\foot{The derivation of this formula is
just as in the three-fold
case; see \ref\cand{P. Candela and X. C. de la Ossa,
``Moduli Space Of Calabi-Yau Manifolds,'' Nucl.Phys. {\bf B355} (1991)
455.} for an exposition.}
\eqn\nokil{K=-\ln \int_Y\Omega\wedge \bar\Omega.}
We want to analyze a possible singularity of this integral near $z_a=0$
in the limit that the $t_i$ go to zero.  If and only if there is
such a singularity, the distance to $t_i=0$ will be infinite in
the metric \okil.
For analyzing this question, the large $z_a$ behavior, which depends
on how the singularity is embedded in a compact variety $Y$, is immaterial
(as long as there is some cutoff to avoid a divergence at large $z_a$);
we can, for instance, replace $Y$ by the hypersurface in \pxol\ and
restrict the integral to the region $|z_a|<1$.

 To determine the small $z_a$ behavior of the integral, we use a simple
 scaling.
Under $z_a\to \lambda^{r_a}z_a$, $\Omega$ scales like $\lambda^{\Sigma_ar_a-1}$
and so the integral in \nokil\  scales like
$|\lambda|^{2\Sigma_ar_a -2}$.  Small $z_a$ corresponds to small $\lambda$.
Hence the condition that the integral
converges at small $z_a$ is
\eqn\juunico{\sum_ar_a-1>0.}
This is a satisfying result, in that this is the same condition that
was needed to get an $R$-symmetry with positive charges
and to apply the Tian-Yau theorem on existence of asymptotically
cone-like Calabi-Yau metrics.

We will now apply this kind of reasoning to address the following
question, whose importance will become clear:
As $Y$ becomes singular, what is the Hodge type of the
``vanishing cohomology,'' that is, of the part of the cohomology that
``disappears'' at the singularity?  We only have to look at middle
dimensional cohomology, because the deformation of a hypersurface
singularity has cohomology only in the middle dimension.

First let us ask if there is vanishing cohomology of type $(4,0)$.
For this, we normalize the holomorphic $(4,0)$-form $\Omega$ of $Y$ in
such a way that far from $z_a=0$ it has a limit as $t_i\to 0$.
Then we ask if the integral
\eqn\omop{\int_Y \Omega\wedge \bar\Omega}
converges as $t_i\to 0$.
If the answer is {\it no}, then to make the integral converge
as $t_i\to 0$, we would have to rescale $\Omega$ so that in the limit
it vanishes pointwise away from the singularity.  Then in the limit
$t_i\to 0$, $\Omega$ would be a closed four-form that is non-zero but
vanishes away from the singularity.
There would thus be vanishing cohomology of type
$(4,0)$.  If the answer is {\it yes}, there is no vanishing cohomology
of type $(4,0)$.

We have already seen that convergence of the integral in \omop\ is
the condition that the singularity is at finite distance in moduli
space.  Hence, singularities that can arise in the dynamics of a compact
Calabi-Yau four-fold have no
vanishing cohomology of type $(4,0)$.

Now let us look for vanishing cohomology of type $(3,1)$.
The $(3,1)$ cohomology is generated by $\Omega_i=D\Omega/Dt_i$, where
$D/Dt_i$ is the covariant derivative computed using the Gauss-Manin
connection.  To determine if $\Omega_i$ is a vanishing cycle,
we need to examine the integral
\eqn\kilmo{\int_Y\Omega_i\wedge \bar\Omega_i,}
and ask if it is finite as all $t_j\to 0$.  If not, then to make
the integral converge, we would have to rescale $\Omega_i$ by a function
of the $t_j$, and in the limit $t_j\to 0$,
$\Omega_i$ would represent a nonzero $(3,1)$
cohomology class that vanishes away from the singularity, or in other
words a piece of the vanishing cohomology of type $(3,1)$.
The integral \kilmo\
is more conveniently written as
\eqn\nilmo{{\partial^2\over \partial t_i\partial \bar t_i}\int_Y
\Omega\wedge \bar\Omega.}
Whether this integral converges can, again,    be determined by
scaling.  If the function $A_i$ in \pxol\ scales under $z_a\to \lambda^{r_a}
z_a$ as $\lambda^{s_i}$, then $t_i$ scales like $\lambda^{1-s_i}$ and \nilmo\
scales like $|\lambda|^{w_i}$ with
\eqn\itsw{w_i= 2\left(\sum_ar_a+s_i-2\right).}
Vanishing $(3,1)$ cohomology arises when $w_i\leq 0$, so that the integral
in \nilmo\ diverges near $z=0$.
The most dangerous case is for $A_i=1$, $s_i=0$.
The condition that $w_i>0$ for all $i$, so that   there is no
vanishing $(3,1)$ cohomology, is thus
\eqn\bitsw{\sum_a r_a >2.}

We can classify the models that obey this condition.  Consider
a Landau-Ginzburg model with chiral superfields $\Phi_a$, $a=1,\dots, 5$
and superpotential $F(\Phi_1,\dots,\Phi_5)$.  If $\Phi_a$ have degree
$r_a$ and $F$ has degree one, then the central
charge of this model is $\hat c=\sum_{a=1}^5(1-2r_a)=5-2\sum_ar_a$.
The condition \bitsw\ thus amounts to\foot{Note that in terms
of $ \hat c$ the condition that the local
singularity of the fourfold be at finite distance in moduli space
\juunico\ is that
$\hat c<3$, which generalizes for an $n$-fold singularity
to $\hat c<n-1$.}
\eqn\nitsw{\hat c<1.}
The singularities that obey this condition are the A-D-E singularities.
They are given, in a suitable set of coordinates, by
\eqn\noxxy{F(z_1,\dots,z_5)=H(z_1,z_2,z_3)+z_4^2+z_5^2,}
where $H(z_1,z_2,z_3)=0$ is the equation of an A-D-E surface singularity.

\bigskip\noindent{\it Application To Hypersurface}

We have developed this discussion for the case of a compact
Calabi-Yau manifold $Y$ that develops a hypersurface singularity,
but it is more in the spirit of the present paper to decompactify $Y$
and focus
on the hypersurface itself, that is to consider
$M$-theory on $\R^3\times X$, where $X$ is a hypersurface that develops
the given singularity.  This
is the natural framework for studying $M$-theory near a singularity,
with extraneous degrees of freedom decoupled.
Let us therefore now explain the significance of the above results
for this case.

If we work on the noncompact hypersurface, the condition
that $\sum_ar_a+s_i>2$, which ensures that there is {\it not}
a divergence of $\int|\Omega_i|^2$ near $z_a=0$, also ensures that there
{\it is} such a divergence near $z_a=\infty$.
The large $z_a$ divergence means that, in $M$-theory on $\R^3\times X$,
the modes that deform the singularity of $X$ have divergent kinetic
energy and are not dynamical.  They correspond, instead, to coupling
constants of the theory near the singularity; they can be specified
externally as part of the definition of the problem.

In the A-D-E examples, the complex structure modes are all non-dynamical
in this sense.
For other examples,  positivity of \itsw\ does not hold for all $i$,
and therefore some of the complex structure deformations of $X$ are
dynamical; they vary quantum mechanically in the theory at the singularity.
Only those modes for which $w_i>0$ can be specified externally and
represent coupling constants.

Now let us consider the Hodge type of the $G$-field in the hypersurface
case.  For unbroken supersymmetry in flat spacetime,
$G$ must be a harmonic ${\bf L}^2$ form
of type $(2,2)$ \beckers.  It must, as well, be integral and ``primitive.''

For hypersurface singularities with asymptotically conical
metrics of the type predicted
by the Tian-Yau theorem, the condition that  $G$ be a harmonic ${\bf L}^2$
form
is a mild one in the following sense.  For an asymptotically conical
metric on a manifold $X$, one expects the
space of ${\bf L}^2$ harmonic forms of degree $i$ to be
isomorphic to the image of $H^i_{cpct}(X;{\bf R})$ in $H^i(X;{\bf R})$.
For hypersurface
singularities of complex dimension four,  there is only four-dimensional
cohomology, so we expect ${\bf L}^2$ harmonic forms of degree four
only.  Assuming there are no
null vectors in $H^4_{cpct}(X)$, the image of $H^i_{cpct}(X;{\bf R})$
in $H^i(X;{\bf R})$ is all of $H^i(X;{\bf R})$, so one expects that
all of the
four-dimensional cohomology is realized by ${\bf L}^2$ harmonic forms.

What about the requirement that $G$ be primitive?
Primitiveness means that $K\wedge G=0$, where $K$ is the K\"ahler form.
If $G$ is an ${\bf L}^2$ harmonic four-form on a manifold whose
${\bf L}^2$ harmonic forms are all four-forms, then $K\wedge G$ is
automatically zero (if not zero, it would be an ${\bf L}^2$ harmonic
six-form).
  Thus, for singularities of this type,
the condition that $G$ should be primitive is automatically obeyed.
\foot{A different explanation of this is as follows. In section 2.2,
we compactified $X$ to a complete but non-Calabi-Yau variety $Y'$ by
adding a divisor $D$ at infinity.  $D$ is an ample divisor, and
the ``primitive'' cohomology in this situation is the cohomology
that vanishes when restricted to $D$.  This is certainly so for the
vanishing cohomology, whose support is far from $D$.}
In section 5, we will examine a singularity of a different
sort  for which primitiveness
of $G$ is an important constraint.

The remaining constraint that we have not examined yet is a severe
constraint in the case of hypersurface singularities.
This is the condition that $G$ should be of type $(2,2)$.
For  A-D-E singularities, as we have seen above, the vanishing cohomology
is all of type (2,2), so the ${\bf L}^2 $ harmonic forms have this
property.  For other singularities, with $\sum_ar_a<2$, there  is
vanishing cohomology of types (3,1), (2,2), and (1,3).  Under such
conditions, it is generically very hard to find a non-zero four-form
that is of type
(2,2) and integral.  Once an integral four-form $G$ is picked, requiring
that it be of type $(2,2)$ will generally put restrictions on the complex
structure of $X$.  Since some of the complex structure modes are
dynamical whenever there is vanishing (3,1) cohomology, the restriction
on complex structure
that is entailed in making $G$ be of type (2,2)
is likely to play an important role in the dynamics of these models.
In this paper,
to avoid having to deal with the dynamical complex structure modes
and the Hodge structure of the singularity,
we will study in detail only the A-D-E singularities. (For fourfold
 examples where
moduli are dynamically frozen see \ref\lerc{W. Lerche,
``Fayet-Iliopoulos Potentials From Four-Folds,'' JHEP {\bf 9711:004}
(1997)}.)

Here is another way to see  the distinguished nature of the
A-D-E singularities.  As we explain in Appendix I, the intersection form
on $H^4(X,\Z)$ is positive definite on the primitive cohomology
of type (2,2), and negative definite on the primitive cohomology
of types (3,1) and (1,3).  Hence, in particular, having the primitive
cohomology be entirely of type (2,2) is equivalent to positive
definiteness of the intersection form on $H^4(X;\Z)$.  For an intersection
form specified by a bouquet of spheres to be positive definite
is a condition that singles out the A-D-E Dynkin diagrams, so again
we see that the A-D-E singularities are the ones with vanishing
cohomology that is entirely of type (2,2).

\subsec{Interpretation Of Constraints On $G$}

Since the constraints on $G$ found in \beckers\ have played an important
role in this discussion, we will pause here to attempt to gain
a better understanding of these constraints.

We consider compactification of $M$-theory on a compact four-fold $Y$.
We first suppose that $G$ is zero.  Variations of the Calabi-Yau metric
of $Y$ arise either from variations of the complex structure or
variations of the K\"ahler structure.
The variations of the complex structure are parametrized classically
by complex parameters $t_i$, which we promote to chiral superfields
$T_i$.  If $h^{p,q}$ is the dimension of the Hodge group $H^{p,q}(Y)$,
then the number of $T_i$ is $h^{3,1}$.  The K\"ahler structure
is parametrized classically by $h^{1,1}$ real parameters $k_i$.
Compactification of the $C$-field on $Y$ gives rise to $h^{1,1}$
$U(1)$ gauge fields $a_i$ on $\R^3$ whose duals are scalars $\phi_i$
that combine with the $k_i$ to make $h^{1,1}$ chiral superfields
that we may call $K_i$.

If $G=0$, the expectation values of the $T_i$ and $K_i$ are arbitrary,
in the supergravity approximation to $M$-theory.  (Instantons can lift
this degeneracy \ref\ugwitten{E. Witten,``Non-perturbative Superpotentials
in String Theory,'' Nucl. Phys. {\bf B474} (1996) 343.}.)
For non-zero $ G$,  this is not so.
After picking an integral four-form $G$ (which must be such that
$\int_X G\wedge G>0$ or the equations we are about to write will
have no solutions), we must adjust the
complex structure of $X$ so that
\eqn\juxin{G_{0,4}=G_{1,3}=0,}
and the K\"ahler structure of $X$ so that
\eqn\lully{G\wedge K = 0.}
In \juxin, $G_{p,q}$ denotes the $(p,q)$ part of $G$.

We want to describe an effective action for the $T_i$ and $K_j$
that accounts for \juxin\ and \lully.  Since supersymmetric actions of
the general kind $\int d^4\theta(\dots)$ do not lift vacuum degeneracies,
we look for F-term interactions.  Thus, we want a superpotential
$W(T_i)$ that will account for \juxin, and an analog of
a superpotential $\tilde W(K_j)$ that will account for \lully.
In three dimensions, the fields $K_j$ are in vector
multiplets, and the function $\tilde W(K_j)$ is related
by supersymmetry to Chern-Simons couplings for those multiplets.
We will not try to work out the full details of this here.
Upon dimensional reduction to two dimensions, the $K_j$
become twisted chiral multiplets and $\tilde W(K_j)$ becomes
the twisted chiral superpotential. Therefore, we will
somewhen loosely call $\tilde W$ a superpotential.

To obtain \juxin, we propose to let $\Omega$ be a holomorphic four-form
on $Y$, and take
\eqn\xonn{W(T_i)={1\over 2\pi}\int_Y\Omega\wedge G.}
This object is not, strictly speaking, a function of the $T_i$ but
a section of a line bundle over the moduli space ${\cal M}$ of
complex structures on $Y$ (on which the $T_i$ are coordinates),
since it is proportional
to the choice of $\Omega$.
Let ${\cal L}$ be the line bundle over ${\cal M}$
 whose fiber is the space of holomorphic four-forms on $Y$.
The K\"ahler form of ${\cal M}$ can be written
\eqn\nonc{\omega =-\partial\bar\partial {\rm ln}\int_Y\Omega\wedge
\bar\Omega,}
in other words it
is $\partial\bar\partial \ln |\Omega|^2$ for $\Omega$
any section of ${\cal L}$,
and this means \ref\bagwit{J. Bagger and E. Witten,``Quantization
of Newton's Constant in Certain Supergravity Theories,''
Phys. Lett. {\bf B115} (1982) 202.}
that $W$ should be a section of ${\cal L}$.  Thus, the linear
dependence on $\Omega$ in \xonn\ is the right behavior of a superpotential.
In supergravity with four supercharges,
the condition for unbroken supersymmetry in flat space is $W=dW=0$.
With $W$ as in \xonn, the condition $W=0$ is that $G_{0,4}=0$.
Also, since the objects $d\Omega/dt_i$ generate $H^{3,1}(Y)$,
the condition $dW=0$ is that $G_{1,3}=0$.
So we have found the supersymmetric interaction that accounts
for \juxin.

Another way to justify \xonn\ is to consider supersymmetric
domain walls.  The tension of a domain wall obtained by wrapping
a brane on a four-cycle $S$ is the absolute value of $\int_S\Omega$.
If $G$ changes from $G_1$ to $G_2$ in crossing the domain wall,
then $G_2-G_1=2\pi [S]$, so this integral is
\eqn\ucu{{1\over 2\pi}\int_X\Omega\wedge (G_2-G_1).}
In a theory with four supercharges, the tension of a supersymmetric
domain wall is the absolute value of the change in the superpotential
$W$.  So \ucu\ should be the change in $W$ in crossing the domain
wall, a statement that is clearly compatible with \xonn.

In a similar spirit, one can readily guess the interaction responsible
for \lully:
\eqn\kxob{\tilde W(K_i)=\int_X{\cal K}\wedge {\cal K}\wedge G.}
Here ${\cal K}$ is a complexified K\"ahler class whose real
part is the ordinary K\"ahler class $K$.  The condition $d\tilde W=0$
is ${\cal K}\wedge G=0$, whose real part is \lully.  $\tilde W=0$ is
a consequence of this, and imposes no further condition.

In $M$-theory on a {\it compact} Calabi-Yau four-fold $Y$,
near a hypersurface singularity, the relation of the change in the
superpotential in crossing a domain wall to \ucu\ shows that
$W$ cannot vanish in all vacua.
In the first version of the present paper it was conjectured
that vacua with non-zero $W$
correspond to supersymmetric AdS compactifications.
However, a more careful analysis in the revised Appendix II
(triggered by comments of J.~Polchinski)
shows that four-fold compactifications with $G_{0,4} \neq 0$
lead to non-supersymmetric theories which are classically
scale invariant\foot{In general, quantum corrections
presumably break the classical scale invariance which
changes the normalization of the Lagrangian.
For instance, five-brane instantons
are expected to modify the effective superpotential and cause
the model to roll down to a supersymmetric vacuum with negative
cosmological constant.}, the so-called ``no-scale models''.
In fact, since the supersymmetry conditions \juxin\ and \lully\
are invariant under overall rescaling of the metric on $Y$,
this conclusion is needed for the superpotential
$W(T_i,K_j)=W(T_i)+\widetilde W(K_j)$
to describe correctly the effective dynamics of M-theory
on a Calabi-Yau four-fold with a $G$-flux.
Going to a non-compact manifold has the effect of decoupling gravity
and allows us to avoid this problem in the present discussion.

Going back to supersymmetric compactifications to $\R^3$,
it is interesting to compactify one
of the directions in $\R^3$ on a circle and consider Type IIA on
$\R^2 \times Y$.  The above analysis carries over immediately for
supersymmetric
vacua with a nonzero value of the
Ramond-Ramond four-form $G$.  However, in Type IIA string theory,
 in view of
mirror symmetry and other $T$-dualities, one naturally thinks that
one should construct a more general effective superpotential to incorporate
the possibility of turning on a full set of Ramond-Ramond fields,
and not just the   four-form.  Indeed, the mirror of $G_{0,4}$ would
be the RR zero-form (responsible \ref\strompol{J. Polchinski
and A. Strominger,``New Vacua for Type II String Theory,''
Phys. Lett. {\bf B388} (1996) 736.}
for the massive deformation of Type IIA supergravity), and the mirror
of $G_{1,3}$ would be the RR two-form.
This is under investigation \ref\sguk{S.~Gukov, ``Solitons, Superpotentials
and Calibrations", Nucl.Phys. {\bf B574} (2000) 169, hep-th/9911011.}.

\bigskip\noindent
{\it Physical Interpretation}

We will now discuss the physical
interpretation of the superpotentials that we have computed.

We have computed the superpotential as a function of the superfields
$T_i$ and $K_j$ with all other degrees of freedom integrated out.
For $Y$ a large, smooth Calabi-Yau four-fold, this is a very natural
thing to do, since the superfields $T_i$ and $K_j$ are massless if
$G=0$, while other superfields are massive.
However, we have argued that as one approaches a singularity,
there are different vacuum states in the theory at the singularity
that are specified by different choices of the $G$-field.  We will
interpret the theory near the singularity as a theory of dynamical
chiral fields $\Phi_\alpha$ such that the critical points of the superpotential
as a function of $\Phi_\alpha$ are given by the possible choices of $G$-field.
Thus, a more complete description of the theory would involve
a superpotential function $\widehat W(\Phi_\alpha;T_i,K_j)$, such
that the function $W(T_i,K_j)=W(T_i)+\widetilde W(K_j)$ is obtained by
extremizing $\widehat W$ with respect to the $\Phi_\alpha$.
For fixed choices of $T_i$ and $K_j$, the extremization with respect
to $\Phi_\alpha$ has different solutions, corresponding to the different
choices of $G$.
It is very difficult to see the superfields $\Phi_\alpha$ explicitly,
but for suitable examples
 we will identify the superpotential function $\widehat W(\Phi_\alpha;
T_i,K_j)$ in section 3 by studying the soliton structure.

\subsec{Analogs For Type IIA And $F$-Theory}

We have formulated the discussion so far in terms of
$M$-theory on $\R^3\times Y$, with $Y$ a Calabi-Yau four-fold,
but there are close analogs for Type IIA
on $\R^2\times Y$ and (if $Y$ is elliptically fibered)
for $F$-theory on $\R^4\times Y$.

The analysis of \beckers\ carries over to Type IIA, with
$G$ now understood as the Ramond-Ramond four-form field.
Our analysis of the vacuum structure also carries over readily to this
case.  One obvious change is that domain walls are now constructed
from four-branes (with world-volume $\R\times S\subset \R^2\times Y$).
Another obvious change is
that, in Type IIA,
the space-filling membranes that contribute to the formula \nuffy\
for the flux at infinity are replaced by space-filling fundamental strings.
Also, in the Type IIA case, alongside the Ramond-Ramond four-form,
one would want to incorporate the Ramond-Ramond zero-form and two-form,
as we have discussed briefly in section 2.4.

In going to  $F$-theory,
the space-filling membranes that contribute to the flux $\Phi$ at
infinity are replaced by space-filling threebranes.  Also
we need to discuss the $F$-theory analog of the $G$-field.
Let $Y$ be a four-fold that is elliptically fibered over a base $B$.
Let $\theta^i$, $i=1,2$,
be a basis of integral harmonic one-forms on the fibers, and
let $\chi$ be an integral two-form generating the two-dimensional
cohomology of the fibers.  Then a four-form $G$ on $Y$ has at the level
of cohomology an expansion
\eqn\ucu{G=g+p\wedge \chi +\sum_i H_i\wedge \theta^i,}
where $g,p$, and $H_i$ are respectively forms of degree $4,$ $2$,
and $3$ on $B$.  ($H_i$ is a three-form on $B$ with values in the
one-dimensional cohomology of the fibers, while $g$ and $p$ are ordinary
four- and two-forms on $B$.)
If $G$ is primitive, then it is in particular self-dual (see
Appendix I).  For $G$ to be integral, $g$ and $p$ must be integral.
Self-duality of $G$ gives a relation between $g$ and $p$
which, in the limit that the area of the fibers of $Y\to B$ is
very small, is impossible to obey if $g$ and $p$ are non-zero and integral.
Hence, the surviving part of $G$ in the $F$-theory limit is
contained in the $H_i$, which are interpreted physically as the Neveu-Schwarz
and Ramond-Ramond three-form field strengths of Type IIB superstrings.
With $g=p=0$, $G$ is an element of the primitive cohomology of $Y$
that is odd under the involution that acts as $-1$ on the elliptic
fibers and trivially on the base.

In terms of a Type IIB description, we have the following structure.
Let $H^{NS},H^{R}$ denote the NS and Ramond
three-form field strengths.
 Let $B$ denote the base
of F-theory ``visible'' to type IIB.  Consider
$$H^+=H^R-\tau H^{NS}$$
$$H^-=H^R-{\bar \tau}H^{NS}$$
We view $\tau$ as varying over $B$ with monodromies
around the loci of seven-branes by $SL(2,{\bf Z})$ transformations
$$\tau \rightarrow {a\tau +b\over c\tau +d}$$
Under such transformations
$$H^+\rightarrow (c\tau +d)^{-1} H^+$$
$$H^-\rightarrow (c{\bar \tau}+d)^{-1} H^-$$
A supersymmetric configuration in this context is obtained by choosing
an integral $(1,2)$ form on the base, $H$, well defined modulo transformation
by $(c\tau +d)^{-1}$ around the 7-branes. Alternatively, $H$ is
a section of $\Omega^{1,2}\otimes {\cal L}$ where ${\cal L}$
is a line bundle
over $B$ whose first chern class is $c_1({\cal L})=-12 c_1(B)$.  Then we
identify
$$H^+=H ,\qquad H^-={\overline H}$$
Moreover we require that $H\wedge k=0$ where $k$ denotes the
K\"ahler class of $B$.  In this case a given model is specified by fixing
$$\Phi=N+{1\over 4\pi ^2}\int_B {1\over \tau_2} H\wedge {\overline H}$$
where $N$ denotes the number of $D3$ branes.

To describe the domain walls, recall that
 one can interpret $F$-theory on $\R^4\times Y$ in terms of
Type IIB on $\R^4\times B$ with $(p,q)$-sevenbranes on a certain
locus  $L\subset B$. Domain walls across which the $H_i$ jump are described by
a five-brane wrapped on $\R^3\times V\subset \R^4\times B$ with
$V$ a three-cycle in $B$.  The $(p,q)$ type of the five-brane varies
as $V$ wraps around the discriminant locus in $B$.

\def\C{{\bf C}}
This is a rather complicated structure in general, but to
study the local behavior near a singularity, it simplifies
considerably.  One reason for this is that near an isolated singularity,
one can replace $B$ by $\C^3$.  If we pick coordinates $z_1,z_2,z_3$
on $\C^3$, then the elliptic fibration over $\C^3$ can be described
very explicitly by a Weierstrass equation for additional complex
variables $x,y$:
\eqn\ikkop{y^2=x^3
+f(z_1,z_2,z_3)x+g(z_1,z_2,z_3).}
The fibers degenerate over a singular locus $L$ which is the
discriminant of the cubic, given by $\Delta=0$, where
\eqn\kikkop{\Delta= 4f^3(z_1,z_2,z_3)+27g^2(z_1,z_2,z_3).}
A singular behavior of the elliptic fibration $Y$ just corresponds
in this language to a singularity of the hypersurface $L\subset \C^3$.
We are interested in a singular point of $L$ at which
$4f^3+27g^2=0$ but $f$ and $g$ are not both zero.\foot{
Singularities with $f$ and $g$ both zero are composite 7-branes
of various types (the order of vanishing of discriminant would
be bigger than 1).  For such cases the simplifications described
in the text do not arise and the full structure of $(p,q)$ sevenbranes
is relevant.}  Near such a singular
point, the detailed construction of $\Delta$ in terms of $f$ and $g$
is irrelevant, and and one can regard $L$ as a fairly generic
deformation of a hypersurface singularity $\Delta=0$.

Actually, the full structure of $(p,q)$ sevenbranes also simplifies
in this situation.  The deformation of an isolated
surface singularity is topologically a bouquet of two-spheres,
and in particular simply-connected.  Hence, there is no
monodromy around which the type of brane can change; the $(p,q)$ type of
the sevenbrane is fixed, and one can think of it (for example)
as a D7-brane.  Thus, the $F$-theory analog of a Calabi-Yau fourfold
singularity is a more elementary-sounding problem: the study
of a D7-brane in $\R^{10}=\R^4\times \C^3$
whose worldvolume is $\R^4\times L$, where $L\subset \C^3$
is developing a singularity.

Now, let us describe the vacuum states and domain walls in this
context.  The D7-brane supports a $U(1)$ gauge field, whose
first Chern class is an element of $H^2(L;\Z)$.  This
group is a lattice $\Gamma^*$, whose rank is the number of
two-spheres in the bouquet.  A D5-brane can end on a D7-brane;
its boundary couples magnetically to the gauge field on the D7-brane.
Hence the domain walls across which the first Chern class jumps
are built from fivebranes of topology $\R^3\times V$, where
$V$ is a three-manifold in $\C^3$ whose boundary lies in $L$.
In crossing such a domain wall, the first Chern class jumps
by the cohomology class $[\partial V]$, which is an element
of $\Gamma= H_2(L;\Z)=H^2_{cpct}(L;\Z)$.
Poincar\'e duality for noncompact manifolds asserts that $\Gamma$ and
$\Gamma^*$ are dual, and\foot{If there are null vectors in $\Gamma$,
there is a slightly more elaborate story as mentioned in section 2.2.}
the natural map $i:H^2_{cpct}(L;\Z)\to
H^2(L;\Z)$ gives an embedding of $\Gamma$ in $\Gamma^*$.
Thus we have a familiar
situation: the vacuum is determined by a point in a lattice $\Gamma^*$,
and in crossing a domain wall it can jump by an element of
a sublattice $\Gamma$.  $\Gamma$ is endowed with an integral
quadratic form (the intersection pairing), and as the notation
suggests, $\Gamma^*$ is the
dual lattice of $\Gamma$ with respect to this pairing.

The A-D-E singularities will furnish important examples in the
present paper, for reasons that we have already explained.  Thus,
let us explain how they arise in the $F$-theory context.
An example of an elliptic four-fold fibration $Y$ with an isolated singularity
is given by the following Weierstrass
equation:
\eqn\normoo{y^2=x^3-3a^2 x+(H(z_1,z_2,z_3) +2a^3).}
Here $a$ is an arbitrary non-zero constant, and $H$ is a quasihomogeneous
polynomial
describing a singularity in three variables at $z_1=z_2=z_3=0$.
If we shift $x$ to $x+a$, the equation becomes
\eqn\tormoon{y^2=x^3+3ax^2 +H(z_1,z_2,z_3),}
and this makes it obvious that the singularity of the elliptic
fibration is obtained by ``stabilizing'' the surface singularity
$H=0$ by adding the quadratic terms $3a^2x^2-y^2$ (the $x^3$
term is irrelevant near the singularity, which is at $x=y=0$).
The equation for the discriminant locus $L\subset {\bf C}^3$ reduces to $H=0$
(plus higher order terms that are irrelevant near the singularity).
So the singularity of the elliptic four-fold is just the
``stabilization'' of the singularity $L$.
To obtain the A-D-E singularities, for both the surface $L$
and the four-fold $Y$,
we need only select the appropriate $H$:

\eqn\adesingularities{\eqalign{H&=z_1^n+z_2^2+z_3^2 ~~\qquad A_{n-1}\cr
H&=z_1^n+z_1z_2^2+z_3^2\qquad D_{n+1}\cr
H&=z_1^3+z_2^4+z_3^2 ~~\qquad E_6\cr
H&=z_1^3+z_1z_2^3+z_3^2 \qquad E_7\cr
H&=z_1^3+z_2^5+z_3^2 ~~\qquad E_8.\cr}}
%
%
%

\subsec{Conformal Field Theory: First Results}

Given Type IIA, $M$-theory, or $F$-theory on a singular geometry,
one natural question is whether a non-trivial conformal field theory
arises in the infrared.

In one situation, an affirmative answer to this
question is strongly
suggested by recent literature.  This is the case of $M$-theory
at a quasihomogeneous four-fold singularity (for the present discussion
this need {\it not} be a hypersurface singularity) with a large
value of the conserved quantity $\Phi$ that was introduced
in section 2.2:
\eqn\hucup{\Phi=N+{1\over 2}\int_X{G\wedge G\over (2\pi)^2}.}
We suppose that the four-fold $X$ is a cone over a seven-manifold
$Q$.  Consider $M$-theory on $\R^3\times X$, with a specified
(flat) $C$-field at infinity that we call $C_\infty$,
and with a very large number
of membranes near the singularity, such that the total membrane
charge (including the contribution of the $C$-field) is $\Phi$.
\nref\keh{A. Kehagias, ``New Type IIB Vacua And Their $F$-Theory
Interpretation,'' hep-th/9805131.}
\nref\kleb{I. Klebanov and E. Witten,``Superconformal Field Theory
on Three-Branes at a Calabi-Yau Singularity,'' Nucl. Phys. {\bf B536}
(1998) 199.}
\nref\morrison{D. Morrison and R. Plesser,``Nonspherical Horizons,''
hep-th/9810201.}
This system is believed \refs{\keh - \morrison}
to be described in the infrared by
a conformal field theory that is dual to $M$-theory on ${\rm AdS}_4\times Q$,
with a constant curvature (but topologically trivial)
$C$-field on ${\rm AdS}_4$ that depends on $\Phi$,
and a flat but topologically nontrivial $C$-field on $Q$ that is equal to
$C_\infty$.
\nref\sethitp{S. Sethi, ``A Relation Between $N=8$ Gauge Theories
In Three Dimensions,'' JHEP {\bf 9811:003, 1998}, hep-th/9809162.}%
For a special case in which the role of $C_\infty$ has been analyzed
(for $Q={\bf RP}^7$) see \sethitp.

The ${\rm AdS}_4$ dual of this CFT depends only
on what one can measure on $Q$, that is
  $C_\infty$ and $\Phi$, and not  the detailed way of decomposing
$\Phi$ in terms of $N$ and $G$ as in \hucup.  That decomposition
arises if one makes a deformation of the theory, deforming
$X$ to a smooth hypersurface.  $M$-theory on $\R^3\times X$ with
$X$ such a smooth hypersurface has vacua corresponding to all
choices of $N$ and $G$ obeying \hucup.  When $X$ develops a singularity,
the $G$-field apparently
``disappears'' at the singularity, and the decomposition
of $\Phi$ into membrane and $G$-field terms is lost.

Note that the vacua with $N\not= 0$ do not have a mass gap even after
deforming to smooth $X$.
There are at least massless modes associated with the motion
of the membranes on $X$.  To get a theory that
after deformation of the parameters flows in the infrared to massive
vacua only, one must
set $\Phi$ to the smallest possible value for a given value of
$C_\infty$, so that after deforming to a smooth $X$,
$N$ will be zero for all vacua.  We recall that $C_\infty$ determines
a coset in $\Gamma^*/\Gamma$.  To get a massive theory,
$\Phi$ must equal the minimum of
\eqn\kxonp{\half\int_X{G\wedge G\over (2\pi)^2},} with $G$ running
over the elements of the coset of $\Gamma^*/\Gamma$ determined
by $C_\infty$; the massive vacua are in correspondence
with the choices of $G$ that achieve the minimum.

Our goal in the next two sections will be to analyze, for
the A-D-E singularities,  the ``massive'' models
just described.  The analysis
will be made by analyzing the domain wall structure, or, as it
is usually called in  two dimensions, the soliton structure.
To justify the analysis, we need to know that there are no
quantum corrections to the classical geometry (which we will use
to find the solitons).  Such corrections would come from appropriate
instantons.
For example, for Type IIA on a Calabi-Yau threefold
near the conifold singularity,  the Euclidean D2 brane instantons
wrapped around the ${\bf S}^3$ in the conifold smooth out the singular
classical
geometry \nref\stret{K. Becker, M. Becker and A.
Strominger,``Five-brane, Membranes and Nonperturbative String
Theory,'' Nucl. Phys. {\bf B456} (1995) 130.}\nref\ov{H. Ooguri
and C. Vafa,``Summing up D-instantons,'' Phys. Rev. Lett.
{\bf 77} (1996) 3296.}\nref\ss{N. Seiberg and S. Shenker,``
Hypermultiplet Moduli Space and String Compactification
to Three-Dimensions,'' Phys. Lett. {\bf B388} (1996) 521.}%
\refs{\stret - \ss}.
Likewise, in $M$-theory compactifications on
suitable Calabi-Yau four-folds, a
superpotential is generated by wrapped Euclidean fivebranes
 \ugwitten.
Such effects, however, are absent in the examples we are considering.
For example,  in the $F$-theory, we are really studying, as we have
explained above,
a sevenbrane on $L\subset \C^3$.
Since the
${\bf C}^3$  has no non-trivial cycles,
the relevant instantons will have to end on $L$, in order to have
finite action.  For the instanton
to affect the quantum moduli space it has to be BPS, which
in particular requires that the boundary of the instanton
be a non-trivial compact cycle in $L$.  In Type IIB string theory
the only possible candidate instantons which could end on
a sevenbrane are fivebranes and onebranes (of appropriate $(p,q)$ type).
Viewing them as instantons, their boundaries would be
five- and one-dimensional respectively. So if $L$ has no non-trivial compact
five- or one-dimensional cycles, then there are no instantons, and
quantum corrections do not modify the singular classical geometry.
In our case, $L$, whose compact geometry consists of
a bouquet of two-spheres, has only two-cycles,
so there are no instantons.
 This is to be
contrasted with the seemingly similar problem of
$F$-theory on a Calabi-Yau threefold.
In that case, $L$ is a complex curve, with non-trivial one-cycles;
instanton one-branes can and do modify the classical geometry.
This is in fact the $F$-theory version of the description of the corrections
to conifold geometry in Type IIA compactification (and  reduces
to it upon compactification on ${\bf T}^2$).  For $F$-theory on
a four-fold,
 if the singularity of the surface $L$ is
not isolated, then it would generically also
have non-trivial one-cycles and would thus receive corrections.

For $M$-theory or Type IIA near a four-fold hypersurface
singularity, a similar statement holds.
In this case, the local geometry of the deformed
singularity has non-trivial four-cycles only.  Thus there is no
room for instantons, i.e. wrapped Euclidean M2- or M5-branes,
which would require non-trivial three or six-cycles
on $X$.  Thus the classical singularity
survives quantum corrections.

Thus, to analyze the small $\Phi$ theories, we will look
for supersymmetric domain walls using the classical geometry
near the singularity.  The domain walls are constructed from
branes whose volumes vanish as the hypersurface $X$ becomes singular,
so their tensions go to zero.  Thus one can reasonably hope to
get a description in terms of an effective theory that contains
only light degrees of freedom and generates these domain walls.
In fact, for the massive models derived from A-D-E singularities,
we will propose a description in terms of an effective superpotential
for a certain set of chiral superfields that generate the same
soliton structure.  This description will make clear that one should
expect flow to a non-trivial IR conformal field theory in the two-dimensional
cases, and in a few cases in three dimensions.

The basic strategy for identifying a supersymmetric theory
based on its BPS soliton structure
is the classification approach of \ref\ceva{
S. Cecotti and C. Vafa,``On Classification of N=2 Supersymmetric
Theories,'' Commun. Math. Phys. {\bf 158} (1993) 569.}\ to ${\cal N}
=2$ supersymmetric theories
in two dimensions.
Consider a theory with ${\cal N}=2$ supersymmetry
in two dimensions with $k$ vacua, and
consider the integral $k\times k$ matrix $S$ given by
\eqn\kiolo{S=1-A}
where $1$ represents the identity matrix and
$A$ is a strictly upper triangular matrix whose
$A_{ij}$ entry for $i<j$ is the number of nearly massless
 BPS solitons
interpolating between the $i$-th sector and the $j$-th sector
weighted with the index $(-1)^F F$ \ref\neind{S. Cecotti, P. Fendley,
K. Intriligator and C. Vafa,``A New Supersymmetric Index,'' Nucl.
Phys. {\bf 386} (1992) 405.},
 i.e.
$$A_{ij}={\rm Tr}_{ij-solitons}(-1)^FF.$$
It was argued that this
 massive deformation comes from a CFT in the UV limit with central
 charge $\hat c$ and
$k$ chiral fields with $R$-charges $q_i$ which satisfy
$${\rm Eigenvalues}(S^{-t}S)={\rm exp}[2\pi i (q_i-\half {\hat c})]$$
(even the integral part of $q_i$ can be determined from $A_{ij}$ \ceva ).
This is a strong restriction, and in case of deformations
of minimal models, the solitons completely characterize the conformal
theory.  In other words, any theory which upon mass deformation
has the same solitonic structure
as that for a massive deformation of a minimal model
is  equivalent to it!  For non-minimal models, the relation above
between the spectrum of the solitons and the charges of chiral
fields is still a very powerful connection and in particular
fixes the central charge of the corresponding conformal theory.

Above two dimensions, the domain wall or soliton analysis still
identifies an effective superpotential, but it is less common
for a theory with a given superpotential to flow to a nontrivial
IR conformal field theory.  For instance, a theory with a single
chiral superfield $\Phi$ and superpotential $W=\Phi^n$ is believed
to flow to a nontrivial CFT in two dimensions for all $n>2$, while
in three dimensions this  is expected only for $n=3$
\ref\strassler{O. Aharony, A. Hanany, K. Intriligator, N. Seiberg
and M.J. Strassler,``Aspects of N=2 Supersymmetric Gauge Theories
in Three-Dimensions,'' Nucl. Phys. {\bf B499} (1997) 67.},
and  in four dimensions,
it is believed to flow to a trivial IR theory for all $n$.
In any event, our analysis will identify the nonperturbative
massless fields and superpotential
 near the singularity also in three and four dimensions.
Also, even in four dimensions, a $\Phi^n$ superpotential can become
relevant as a perturbation to certain fixed points
\ref\Kut{D. Kutasov,``A Comment on Duality in N=1 Supersymmetric
Non-abelian Gauge Theories,'' Phys. Lett. {\bf B351} (1995) 230.}, so with
some modification of our construction,
the superpotential  we find may eventually be important in analyzing
four dimensional CFT's that arise from string theory.

The soliton analysis will give detailed information about the behavior
for small membrane charge, which is the opposite limit from the AdS description
discussed above
that governs the large charge behavior at least for the $M$-theory
compactifications.  For the Type IIA and $F$-theory compactifications,
the description of the large charge behavior appears to be less simple.

\newsec{Geometry of Domain Walls}

As explained in section 2.6, our task now is
to analyze the soliton structure for certain hypersurface
singularities.  In fact, we will consider the $A_k$ singularities
which were introduced in section 2.2.

Instead of specializing to four-folds, it proves insightful
 to consider the more general problem of identifying BPS states
 of wrapped $n$-branes in a Calabi-Yau        $n$-fold near
 an isolated singularity.
To study the behavior near an $A_k$ singularity, we
consider a local model for a Calabi-Yau $n$-fold given by
$$-P_m(z_1)+z_2^2+...+z_{n+1}^2=0$$
where $P_m(z_1)$ is a polynomial of degree $m=k+1$ in $z_1$.
When $P_m$ has  two equal roots, we get
a singular geometry.  The most singular gemetry arises
when $P_m(z_1)=z_1^m$.  For a generic polynomial
$P_m(z_1)$, the
geometry is not singular and the
compact homology of this manifold has a basis made of $m-1$ spheres
of real dimension $n$ intersecting each other according
to the $A_{m-1}$ Dynkin diagram.  For a particular choice of $P_m$,
we explained how to construct these spheres in section 2.2.
The intersection form on the compact homology is symmetric if
$n$ is even and antisymmetric if $n$ is odd.
We would like
to consider minimal wrapped $n$-branes, i.e. minimal
supersymmetric cycles, in this geometry.  A supersymmetric
cycle is a Lagrangian submanifold (that is, the K\"ahler
form vanishes on it).
Moreover, on a minimal supersymmetric
$n$-cycle the holomorphic $n$-form $\Omega$ of Calabi-Yau
is real (with a suitably chosen overall phase) and gives
the volume of the $n$-brane.  For a minimal supersymmetric
cycle the quantity
$$V=\int_C |\Omega|$$
which is the volume of cycle $C$,
is minimized and is given by
$$V=\alpha \int_C \Omega$$
for some choice of phase $\alpha$. Or stated equivalently, the condition is
that
$$\int_C |\Omega|=\left|\int_C \Omega \right|.$$
which is the condition for minimizing the volume of $C$
 among the Lagrangian submanifolds in a given
homology class.

The holomorphic $n$-form $\Omega$, up to an overall complex
scale factor, is given by
$$\Omega ={dz_1...dz_n\over z_{n+1}}={idz_1...dz_n\over {\sqrt{
z_2^2+...+z_n^2-P_m(z_1)}}}$$
 We would like to minimize the volume form given by $|\Omega|$.
 To count the minimal supersymmetric cycles, we follow
 the strategy in \ref\klmvw{A. Klemm, W. Lerche, P. Mayr,
 C. Vafa and N. Warner,``Self-dual Strings and N=2 Supersymmetric
 Field Theory,'' Nucl. Phys. {\bf B477} (1996) 746.}\ and decompose
 the geometry to the ``fiber and the base'' as follows.
 Suppose $C$ is a supersymmetric minimal cycle.  Consider
 the image of $C$ on $z_1$. This is a one-dimensional
subspace, because for
 a fixed $z_1$,
 the manifold (being defined by $\sum_{j>1}z_j^2=P_m(z_1)$)
 has for its only nontrivial cycle a sphere ${\bf S}_{z_1}^{n-1}$.
 Note that the radius of this sphere is $|P_m(z_1)|^{1/2}$, from
 which one can deduce by scaling that
 \eqn\gumbo{\left| \int_{{\bf S}^{n-1}_{z_1}}{dz_2\wedge \dots \wedge dz_{n-1}
 \over z_n}\right| =|P_m(z_1)|^{(n-2)/2}}
 up to an irrelevant multiplicative constant.
The inverse image of $C$ over a point in the $z_1$ plane
must, if not empty,
 be a minimal cycle, and so must be $n-1$-dimensional;
 hence the image of $C$ in the $z_1$ plane must be one-dimensional.
  The minimization of $ |\Omega| $
 will be done in two steps:  We first consider Lagrangian submanifolds
 $C_f(z_1)$
 for a fixed $z_1$ which minimize the $ \int_{C_f(z_1)} |\Omega|$
and next consider the minimization of the
volume interval over an interval $I$ in $z_1$.
In this way we get using \gumbo\
 \eqn\tomi{\int_{C_f(z_1)\times I} |\Omega|=
\int_I\left|\int_{S^{n-1}(z_1)} \Omega \right|=
\int_I |P_m(z_1)|^{n-2\over 2}dz_1. }
 We now minimize the volume of the supersymmetric
 $n$-cycle with respect to the choice of the
 one-dimensional line segment $I$ representing the image of the
supersymmetric
 cycle on the $z_1$ plane.
 One can allow the line segment to end at some special points on $z_1$
 where $P_m(z_1)=0$, and these are the only allowed boundaries.  In fact,
 precisely if the line segment terminates at zeroes of $P_m$, the
 $D$-brane worldvolume is closed and smooth.
 Indeed the topology of the cycle is an ${\bf S}^n$ which
 can be viewed as an ${\bf S}^{n-1}$ sphere fibered over the interval,
 where at the boundaries of the interval the radius of ${\bf S}^{n-1}$
vanishes.
 The expression \tomi\ is minimized if
 along the segment on $z_1$ plane
 the condition $|P_m(z)|=\alpha P_m(z)$ is satisfied for some
$z$-independent phase
 $\alpha$.  Let us define a function $W$ with
 \eqn\dow{dW=P_m^{n-2\over 2}dz_1}
 In terms of $W$, the condition for minimal volume is
 that the image of the curve
 in the $W$ plane be a straight line along the
 direction specified by $\alpha^{-1}$.  Moreover, the end-points
 of the segment in the $z_1$ plane correspond to critical points in $W$,
 i.e., $dW=0$ (for $n=2$ the endpoints are
 defined by the condition that $P_m(z_1)=0$).
 These conditions are identical \ref\soliton{P. Fendley, S.D. Mathur,
 C.Vafa and N.P. Warner,``Integrable deformations and scattering matrices
 for the N=2 supersymmetric discrete series,'' Phys. Lett. {\bf 243} (1990)
 257.}\ for finding solitons in an ${\cal N}=2$ Landau-Ginzburg theory
 in two  dimension (or more generally,  BPS domain walls
 in theories with four supercharges in dimensions two, three, or four)
 with superpotential given by
 $W$!  If $n$ is even, \dow\
corresponds to a well defined function of
$z_1$.
 If $n$ is odd, it gives rise to a well defined (meromorphic)
 one-form on a hyperelliptic
 cover of the $z_1$ plane, branched over the zeroes of $P_m(z_1)$.

 Strictly speaking we have constructed the supersymmetric cycle
 by assuming that the condition that the cycle $C_f$
 be Lagrangian is the same as being Lagrangian relative to the K\"ahler
 form induced on the fiber. This is not necessarily true.  For
 example if the K\"ahler form has a piece of the form
 $$k=...+f_i dz_1\wedge dz_i^*+...$$
 the condition of Lagrangian gets modified.  In special
 cases, like when the polynomial $P_m$ has real coefficient
 one can use a $Z_2$ antiholomorphic involution to argue
 that the cycles we constructed are both Lagrangian and supersymmetric.
 In the more general case we proceed as follows:  Consider a generic
 $$P_m(z_1)=\prod_i(z_1- a_i)$$
 Consider a one parameter family of Calabi-Yau metrics where
 $$a_i(t)=t a_i.$$
 Note that the BPS states will be the same for all $t$, because
 the condition of the BPS charges getting aligned does
 not change as we change $t$ (the BPS charges only receive
 an overall rescaling).   However to construct the K\"ahler metric
 as a function of $t$ we note that it can be mapped to the previous
 metric by defining
 $${\tilde z_1}= t z_1$$
 $${\tilde z_i}= t^{n/2}z_i\quad for \ i\not=1$$
 Thus we use the $z$ variables but rescale the K\"ahler form accordingly.
 In this way  as $t\rightarrow
 \infty$ the mixed terms in the K\"ahler form are dominated by
 the terms purely in the fiber direction
 (for $n> 2$ which is the case of main interest).
 Therefore in this limit the condition of Lagrangian
 submanifold in the fiber that we have used becomes accurate.

 Let us consider some special cases.  The cases
 for a K3 surface  and for a Calabi-Yau threefold have already been considered
 in \klmvw\ (see also \nref\rabin{J.M. Rabin, Phys. Lett.
 {\bf B411} (1997) 274.}\nref\warns{J. Schulze and N.P. Warner,
 ``BPS Geodesics in N=2 Supersymmetric Yang-Mills Theory,''
 Nucl. Phys. {\bf B498} (1997) 101.}%
 \refs{\rabin,\warns}), and will
be reviewed below.

 \bigskip\noindent
 {\it Solitons for K3}

 In the case $n=2$, the above geometry is the
 complex deformation of the $A_{m-1}$ singularity.
 For any choice of the polynomial $P_m(z_1)$, we expect
 $m(m-1)/2$ solitons (up to the choice of orientation) to
 complete the adjoint representation of
 $U(1)^{m-1}$ to the $SU(m)$ gauge multiplet.  From
 \dow, we see that in this case $W=z_1$. There are $m$ roots
 for $P_m(z)$, and the solitons correspond to straight lines between
 the roots.  Note that this gives $m(m-1)/2$ solitons up to the
 choice of the orientation, as was anticipated.

 \bigskip\noindent
 {\it Solitons for $CY_3$}

 For the case of Calabi-Yau threefolds, $n=3$.  In this case
 $W$ is defined by
 $$dW=P_m(z_1)^{1\over 2} dz_1.$$
 Here $dW$ can be viewed as a meromorphic one-form
 on a hyperelliptic Riemann surface over $z_1$ branched
 over the roots of $P_m(z_1)$.  The geometry of these solitons
 for this class of conformal theories would correspond to
 straight lines on the Jacobian of this surface defined by the
 integrals $dW$ and is presently under study \ref\shva{A. Shapere,
C.Vafa, ``BPS Structure of Argyres-Douglas Superconformal
Theories,"hep-th/9910182.}.

 \bigskip\noindent{\it Solitons for $CY_4$}

 For the case of four-folds, which are of course our main
 focus in the present paper, the definition of $W$ in \dow\ shows
 that $W$ is a polynomial of degree $m+1$ in the $z_1$ plane.
 We have already shown that the conditions for finding the
 solitons in this geometry are the same as those in an LG theory with
 the superpotential $W$.  In this case, however, if we use
 our four-fold in Type IIA superstring theory, the analogy
 becomes more precise:  compactification on the four-fold
 leads to a theory in two dimensions with ${\cal N}=2$, and it is
 natural to identify the corresponding $W$ with the superpotential
 of an ${\cal N}=2$ Landau-Ginzburg theory.  We will indeed
 argue that for a certain choice of the membrane
 charge, the Type IIA on a deformed $A_1$ singularity leads
 to an ${\cal N}=2$ theory with the same $W$ for its superpotential.
   For more general choices of the membrane charge,
 we find closely related Kazama-Suzuki coset models  at level $1$.

 Before we discuss these subtleties, note that even though
 we have $m$ critical points, it is no longer true in general that we
 have $m(m-1)/2$ solitons.  In general the pre-image of a straight
 line connecting the images of critical points
 in the $W$ plane will not connect the critical points in the $z_1$
 space.  In fact as we change the polynomial $P_m(z_1)$, it is
 known that the number of BPS states jumps \neind.
 For some choices of $P_m(z_1)$ we do have exactly the same
 number of solitons as in the K3 case.  For example, it has
 been shown \soliton\ that for
 $$P_m(z_1)=z_1^m-\mu^n$$
 for any constant $\mu$, there is one soliton for each pair
 of $m$ critical points $z_1=\omega \mu $ with $\omega^m=1$,
 though, unlike the K3 case the image in the $z_1$ plane is
 {\it not} a straight line.

It would be interesting to compare the formula for $W$ that we have
deduced from the soliton structure to the analysis of section 2.4.
Although this is guaranteed to work, because both capture the mass
of the BPS soliton, we have not attempted to check this correspondence
explicitly.

\newsec{Identifications With Kazama-Suzuki Models}

Let us consider in more detail Type IIA strings propagating on
a smooth hypersurface $X$ obtained by deforming the $A_{m-1}$ singularity:
$$-P_m(z_1)+z_2^2+z_3^2+z_4^2+z_5^2=0.$$
As explained in section 2, in order to specify the problem fully,
we must fix the value $C_\infty$ of the $C$-field at infinity
and also the flux
\eqn\ruffy{\Phi=N+\half\xi^2 ,}
where $\xi=[G/2\pi]$ is the characteristic class of the $C$-field.

As we explained in section 2.2, $\xi$ is restricted to a fixed coset
in $\Gamma^*/\Gamma$, where $\Gamma$ and $\Gamma^*$ are the root
and weight lattices of the Lie group $SU(m)$.  The coset in which
$\xi$ takes values is determined by $C_\infty$.   For the theory
to have a mass gap, as discussed in section 2.6,
we set $\Phi$ equal to the minimum value of
$\half \xi^2$ (for all $\xi$ in the given coset), so that $N=0$
for all vacua.

This can be made very explicit in the case of the $A_{m-1}$ singularity.
$C_\infty$ takes values in $\Gamma^*/\Gamma$, which is isomorphic
to the center of $SU(m)$, or $\Z_m$.  For $k=0,\dots,m-1$,
if $C_\infty=k$, then to minimize
$\xi^2$, $\xi$ must be a weight
of the $k$-fold antisymmetric tensor product of the fundamental
representation of $SU(m)$.  We denote that representation as $R_k$.
The number of choices of $\xi$ is the dimension of $R_k$
or $m!/k!(m-k)!$.  This is the number of vacuum states of the $k^{th}$
model, if $P_m$ is such that the hypersurface $X$ is smooth.

For $k=0$, there is only one vacuum ($\xi=0$), and the theory is
trivial and massive.
  Let us consider the next simplest case, where $k=1$
  and $\xi$ is a  weight of the fundamental representation of $SU(m)$. In
  this case, we have $m$ vacua. To find the degeneracy
  of the solitons between these vacua, we use the analysis
  of section 3. We found that the solitons are exactly
  the same as those for an ${\cal N}=2$ LG theory with
  a chiral field $\Phi$ and a superpotential
  $W$ obeying $dW/d\Phi=P_m(\Phi)$.  In fact we can identify the
  $m$ vacua with the $m$ critical points of $P_m$, and as we found in
  section 3,
  the condition for the existence of a soliton in the LG theory is
  exactly the same as the condition for a BPS wrapped four-brane in
  Type IIA near the $A_{m-1}$ singularity.  In this case the soliton data
  are enough to determine the theory \ceva\
 as discussed at the end of section 2; the two-dimensional theory
  with superpotential $W$ is the $A_{m}$ minimal model   \refs{\vwa, \mart }.
   So the Type IIA theory near an
  $A_{m-1}$ four-fold  singularity is governed by the $A_{m}$
  ${\cal N}=2$ minimal model.  This model can also
  be viewed as an ${\cal N}=2$ Kazama-Suzuki coset model at level one, of the
form
  $${SU(m)\over SU(m-1)\times U(1)}$$
  For the $M$-theory or $F$-theory near an $A_{m-1}$ four-fold singularity,
  we still get a description in terms of  a chiral field with
  the same superpotential, but in most instances
  (the exception being $\Phi^3$ in three dimensions), a theory
  in three or four dimensions with a $\Phi^{m+1}$ superpotential is
  believed to flow to a free theory in the infrared.

We now consider the other choices of $C_\infty$, so that
$\xi$ is a   weight of the
$k$-fold anti-symmetric product of the fundamental representation
of $SU(m)$ with some $k>1$.
We argue that it has exactly the same solitonic
spectrum as a deformation of the following LG theory, which we will
call the $k$-fold symmetric combination of the $k=1$ model. Consider the
function of $k$ variables
$$W(z_1,\dots, z_k)=z_1^{m+1}+z_2^{m+1}+...+z_{k}^{m+1}.$$
  It is invariant under permutations of the $z_i$,
and so can be expressed as a polynomial in
the elementary symmetric functions
$$x_l=\sum_{i_1<...<i_l} z_{i_1}...z_{i_l}.$$
The superpotential we consider is thus
\eqn\juccy{W(x_1,\dots, x_k)=W(z_1)+\dots +W(z_k).}
The LG model with superpotential \juccy\
has been conjectured in \ref\lvw{W. Lerche, C. Vafa and N.P. Warner,
``Chiral Rings in N=2 Superconformal Theories,'' Nucl. Phys.
{\bf B324} (1989) 427.}\
to be equivalent to the following Kazama-Suzuki coset model at level
1:
$${SU(m)\over SU(m-k)\times SU(k)\times U(1)}$$
For the deformed singularity, with $\partial W=P_m$, we claim
that the deformed LG superpotential is given by
$$W(x_1,...,x_k)=W(z_1)+...+W(z_k)$$
where again what we mean by this expression is that
the superpotential is $W(z_1)+\dots +W(z_k)$ regarded as a polynomial
in the elementary symmetric functions $x_1,\dots,x_k$.
  Let us
see why this LG superpotential has exactly the same solitonic
spectrum that we get for Type IIA at an $A_{m-1}$ four-fold singularity
with $C_\infty=k\,\,{\rm mod}\,\,m$.
It is not too difficult to show  \nref\iint{K. Intriligator,
``Fusion Residues,'' Mod. Phys. Lett. {\bf A6} (1991) 3543.}%
\refs{\ceva,\iint} that
the set of vacua of a LG theory made of a $k$-fold
symmetric combination of a given LG theory (in the sense introduced
above) can be identified with the
$k$-fold {\it antisymmetric} tensor product of the
space of  vacua of the original
LG theory.\foot{The main point that must be shown is that the vacua
in the different factors must be distinct.  To illustrate why, it suffices
to consider the case that $k=2$ and that each individual model
has only one vacuum.  So we start with $m=2$: $W(z)=z^2$.  Then
we write $W(z_1,z_2)=z_1^2+z_2^2$ in terms of the symmetric
functions $x_1=z_1+z_2$, $x_2=z_1z_2$, getting $W(x_1,x_2)=x_1^2-2x_2$.
This function has no critical points, so the combined model has no
supersymmetric vacua, as expected.}
As we already discussed, the $k=1$ model has
a one-variable superpotential $W(z)$, and its
vacua correspond to the fundamental weights of the $SU(m)$ lattice.
Thus we can identify the vacua of the $k$-fold symmetric combination
of the $k=1$ model with the weights of the $k$-fold anti-symmetric tensor
representation $R_k$.
As far as the allowed solitons,
on the LG side, they can be constructed in the decoupled
theory with superpotential $W(z_1,\dots,z_k)=W(z_1)+\dots +W(z_k)$
before re-expressing this in terms of the symmetric functions $x_i$.
In this description, it is clear that soliton states are just
the products of soliton states in the individual one-variable theories,
and that irreducible solitons (which cannot break up into several
widely separated mutually BPS solitons) are solitons in just one of the
variables $z_i$.
So if we label the vacua by
$|i_1,...,i_k\rangle$, with $i_s$ denoting a vacuum in the $s^{th}$
one-particle theory, then the allowed solitons only change
one vacuum index at a time.
So the solitons of a LG theory
that is constructed as a $k$-fold symmetric combination of a
one-variable theory are in natural correspondence with the solitons
of the one-variable theory.
This is the same result that we get from Type IIA
near the $A_{m-1}$ singularity with $C_\infty=k\,\,{\rm mod}\,\,m$.
Indeed, for this model, the solitons are constructed by finding
supersymmetric four-cycles.  The analysis of those cycles in section
3 depends only on the geometry of the hypersurface and
makes no reference to $C_\infty$.  Hence the solitons for any $k$
are in a natural sense the same as the solitons of the $k=1$ model.

In other words the
solitons are in 1-1 correspondence with those roots of $SU(m)$
that appear as solitons for the one-variable LG superpotential given
by $W$.  Whichever roots appear act in the natural way on the weights
of the representation $R_k$.
  In the case
 $W=z^{m+1}-az$,  all the roots appear
with multiplicity 1.  This structure for the solitons
of the deformed Kazama-Suzuki model was suggested in \ref\lewa{W. Lerche
and N.P. Warner,``Polytopes and Solitons in Integrable N=2 Supersymmetric
Landau-Ginzburge Theories,'' Nucl. Phys. {\bf B358} (1991) 571.}\
where it was argued to correspond (with a specific
choice of K\"ahler potential) to an integrable model.

\subsec{Other A-D-E Singularities}

So far we have mainly considered the local singularity to be
$$H(z_1,z_2,z_3)+z_4^2+z_5^2=0$$
with $H$ being an $A_{m-1}$ singularity.  Here we would like
to generalize this to the case where $H$ determines a $D$ or $E$ type
singularity.

The general structure is quite like what we have seen for $A_{m-1}$.
$C_\infty$ takes values in $\Gamma^*/\Gamma$, where $\Gamma$ is the
root lattice of the appropriate simply-connected A-D-E group $G$,
and $\Gamma^*$ is the weight lattice of $G$.  The quotient $\Gamma^*/\Gamma$
is isomorphic to the center of $G$.

Just as in the $SU(m)$ case,
to make possible a deformation to a  massive theory,
we need to pick $\Phi$ so that $\xi$ ranges over the weights of the smallest
representation with a given non-trivial action of the center of $G$.
(If we pick the trivial action of the center, we will get the trivial
representation and  a massive free theory.)
The appropriate representations are the representations with Dynkin label 1.
In the $D_n$ case, the relevant choices
 are the vector and spinor representations.
For $D_{2n}$, there are two different spinor representations, but they
differ by an outer automorphism of $D_{2n}$ and give equivalent theories.
So there are essentially two choices of $C_\infty$ leading to massive
theories based on the $D_n$ singularity.
For the $E_6$ theory, there is only the fundamental 27
dimensional representation (and its conjugate, which gives an
equivalent theory); for $E_7$ there is the fundamental 56 dimensional
representation.  So $E_6$ and $E_7$ lead to one massive theory each.
$E_8$ is simply-connected with trivial center,
so we cannot use it to get a conformal theory with
a massive deformation.

The distinguished representations that we have described are in one-to-one
correspondence with nodes of index 1 on the A-D-E Dynkin diagram
and thence with Hermitian symmetric spaces $G/H$ (where $H$ is obtained
by omitting the given node from the Dynkin diagram).
Apart from the Grassmannians $SU(m)/SU(k)\times SU(m-k)\times U(1)$ that
we have already encountered, these Hermitian symmetric spaces are as
follows.
For the $D_n$ case, there are two inequivalent choices, given by
$${SO(2n)\over SO(2n-2)\times SO(2)} \qquad {\rm for  \ fundamental \ rep.} $$
$${SO(2n)\over U(n)}\qquad \qquad{\rm for\  spinor \ rep.}, $$
and for the $E_6$ and $E_7$ cases one has
$${E_6\over SO(10)\times U(1)}\qquad {\rm for \ fundamental \ rep.}$$
$${E_7\over E_6\times U(1)}\qquad ~~~{\rm for \ fundamental \ rep.} $$
Such a Hermitian symmetric space determines a series of ${\cal N}=2$
Kazama-Suzuki
models (at level $1,2,3\dots$).  It is natural to conjecture that,
as we have found for $A_{m-1}$, the massive models obtained from
Type IIA at an A-D-E singularity are the massive deformations of
the corresponding level 1 Kazama-Suzuki (or KS) models.  As a first check,
it is known that for a level one $G/H$ KS model, the dimension of
the chiral ring is equal to the dimension of the corresponding representation
of $G$.   This in turn is equal to the dimension of the cohomology
of $G/H$ and it was conjectured in \lvw\ that the chiral ring is
isomorphic to the cohomology ring, which in turn was shown to arise
from the ring of an LG theory.  Thus the $G/H$ theories at level
1 were identified with specific LG models.\foot{The higher level
KS models do not generally admit an LG description.}
Moreover, the structure of the solitons for a special (integrable)
deformation of the  KS model at level 1 was studied in \lewa\ and
it was conjectured that the solitons exist precisely for each allowed
single root acting on the corresponding weight diagram.
Though we have not analyzed the BPS spectrum of the D4-branes
in this case to find the multiplicity
of the solitons for each root, it is natural to expect that at least
for specific deformations, just as in the $A_{m-1}$ case, the solitons
are given by the root lattice of the corresponding group with
multiplicity 1.  In this case we would reproduce the solitonic
structure anticipated in \lewa .
 It is quite satisfying
that we apparently get all the Hermitian symmetric space
 KS models at level 1 in
such a uniform way from considering singularities of CY four-folds.

\newsec{Other Types Of Singularity}

The only four-fold singularities that we have so far considered
in any detail are hypersurface singularities.  A Calabi-Yau
four-fold can, however, develop singularities of many different
types.  We cannot offer any sort of overview of the possibilities,
but will briefly analyze two cases in the present section.

\subsec{Hyper-K\"ahler Singularities}

First we will consider what one might call hyper-K\"ahler singularities
-- singularities near which $Y$ admits a hyper-K\"ahler structure,
though $Y$ may not be globally a hyper-K\"ahler manifold.
An important fact here is that $M$-theory compactification on
$\R^3\times Y$ with $Y$ hyper-K\"ahler has ${\cal N}=3$ supersymmetry in
three dimensions, because the space of covariantly constant spinors
on a hyper-K\"ahler eight-manifold is three-dimensional.

To isolate the behavior near the singularity, we replace $Y$ by
an asymptotically conical hyper-K\"ahler manifold $X$ that is developing
a singularity.  We will focus on a
 very concrete example, with  $X=T^*{\bf CP}^2$,
the cotangent bundle of complex projective two-space.  This
hyper-K\"ahler manifold is conveniently obtained by considering
a $U(1)$ gauge theory with eight supercharges, and three
hypermultiplets ${\cal A}^i$, $i=1,2,3$, of charge 1.
\foot{We will present this gauge theory as a formal device, but
it may have a physical interpretation in terms of a membrane probe
of the singularity.}
There is an $SU(3)$ global symmetry group, with the ${\cal A}^i$
transforming as the ${\bf 3}$.
There is also an $SU(2)$ $R$-symmetry group, and it is possible to
add an $SU(2)$ triplet of Fayet-Iliopoulos terms $\vec d$ to the
$D$-flatness equations.
A manifestly $SU(2)$-invariant way to exhibit the $D$-flatness
equations is as follows.  The bosonic parts of the ${\cal A}^i$
can be regarded as a complex field $A^i_\alpha$, $\alpha=1,2$,
transforming as $({\bf 3},{\bf 2})$ under $SU(3)\times SU(2)_R$.
The $D$-flatness condition is
\eqn\juggo{\sum_i A^i_\alpha\bar A_i^\beta
=\vec d\cdot \vec \sigma_\alpha^\beta,}
with $\vec \sigma$ the traceless $2\times 2$ Pauli matrices.
The moduli space $X$ of zero energy states of the classical gauge theory
is the space of solutions of \juggo\ divided by the action of
the gauge theory.
In this description, it is manifest that if $\vec d=0$, then $X$ has
 an $SU(3)\times SU(2)_R$
symmetry, broken if $\vec d\not= 0$ to $SU(3)\times U(1)_R$.  The $SU(3)$
preserves the hyper-K\"ahler structure, and $SU(2)_R$ rotates
the three complex structures on $X$.
If $\vec d=0$, $X$ is a cone over a seven-manifold $Q$ described by
\eqn\hucxxp{\sum_i A^i_{\alpha}\bar A_i^\beta =\delta^\beta_\alpha.}
It is fairly easy to see that this manifold is a copy of $SU(3)/U(1)$,
where the $U(1)$ acts by right multiplication by
\eqn\incon{{\rm diag}(e^{i\theta}
,e^{i\theta},e^{-2i\theta}).}  $SU(3)$ acts on $SU(3)/U(1)$ by left
multiplication,
and $SU(2)_R$ acts by right multiplication by $SU(3)$ elements that commute
with \incon.   Even if $\vec d\not= 0$, $X$ is asymptotic to a cone
over $Q$ at big distances.
The $R$-symmetry
group that acts faithfully on $X$ is actually $SO(3)_R=SU(2)_R/\Z_2$.
That is because the center of $SU(2)_R$ is equivalent to a $U(1)$
gauge transformation.  In $M$-theory on $\R^3\times X$, the three spacetime
supersymmetries transform as a vector of $SO(3)_R$.

Now let us explain why for $\vec d\not=0$, $X$ is equivalent to
$T^*{\bf CP}^2$.
In terms of a description
that makes manifest only half the supersymmetry of the gauge theory,
one can break
up the bosonic part of the ${\cal A}_i$ into pairs of complex fields $B^i$,
$C_i$, transforming as ${\bf 3}$ and $\bar {\bf 3}$ of an $SU(3)$
symmetry group, and with  charges $1$ and $-1$ under the $U(1)$ gauge
group.   (Compared
to the previous description, $B^i=A^i_1$ and $C_i=\bar A_{i\,2}$.)
This description breaks $SO(3)_R$ to $SO(2)_R=U(1)_R$,
with $\vec d$ splitting
as a real component $d_{\bf R}$ and complex component $d_{\bf C}$.
The $D$-flatness equations of the $U(1)$ gauge
theory are in this description
\eqn\humbo{\eqalign{\sum_i|B^i|^2-\sum_j|C_j|^2&=d_{\bf R}\cr
                      \sum_iB^iC_i   & = d_{\bf C}.\cr}}
One must also divide by the action of $U(1)$.
By an $SO(3)_R$ rotation, one can set $d_{\bf C}=0$ and $d_{\bf R}>0$.
The quantities
$\tilde B^i=B^i/\sqrt{d_{\bf R}+\sum_j|C_j|^2}$ obey $\sum_i|\tilde B^i|^2=1$
and, after dividing by the gauge group, define a point in ${\bf CP}^2$.  With
$d_{\bf C}=0$, the second equation in \humbo\ can be interpreted to
mean that $C_i$  lies in the cotangent space to ${\bf CP}^2$, at the
point determined by the $\tilde B^i$.  Thus $X$ is isomorphic
to $T^*{\bf CP}^2$.
  For any manifold $W$, regarded as  the zero section of
$T^*W$, the self-intersection number $W\cdot W$ is equal to the
Euler characteristic of $W$.  The Euler characteristic of ${\bf CP}^2$ is
3, so in our example
\eqn\huvno{W\cdot W=3}
with $W=[{\bf CP}^2]$.

Though turning on $\vec d$ breaks the $SO(3)_R$ symmetry of $X$ to
$SO(2)$, it preserves the hyper-K\"ahler structure and all of the
supersymmetry of $M$-theory on $\R^3\times X$.

The appearance of an $SO(3)_R$ symmetry at $\vec d=0$ is a hint
that $M$-theory on $\R^3\times X$ flows to a superconformal field
theory in the infrared as $\vec d\to 0$.  Indeed, in three spacetime
dimensions with ${\cal N}$ supercharges, the superconformal algebra contains
an $SO({\cal N})_R$ $R$-symmetry group.

To get more insight, let us now analyze the possible $G$-fields on
the smooth manifold $X$ with $\vec d\not= 0$.
Since $X$ is contractible to ${\bf CP}^2$, one has $H^4(X;\R)
=H^4({
\bf CP}^2;\R)$.  The non-zero Betti numbers are $h^0=h^2=h^4=1$.
The cohomology with compact support is, by Poincar\'e duality, the
dual of this, so the non-zero Betti numbers with compact support
are $h^4_{cpct}=h^6_{cpct}=h^8_{cpct}$.
Hence, just on dimensional grounds, the natural map $i:H^k_{cpct}(X;\R)
\to H^k(X;\R)$ is zero except for $k=4$.  For $k=4$, $H^4_{cpct}(X;\R)$
is generated by the class $[W]=[{\bf CP}^2]$, and the nonzero
intersection number \huvno\ implies that $i\not= 0$.  For an asymptotically
conical manifold, one expects the space of ${\bf L}^2$ harmonic
forms to coincide with the image of $i$, so in the present
example we expect precisely one ${\bf L}^2$ harmonic form $\alpha$,
in dimension four.  $\alpha$ is necessarily primitive
with respect to all of the complex structures, since if
$K$ is any of the K\"ahler forms, then $\alpha\wedge K$, if not
zero, would be an ${\bf L}^2$ harmonic six-form.

Hence, turning on a nonzero
$G$-field, proportional to $\alpha$, preserves all of the
supersymmetries.  In fact, we {\it must}
turn on such a $G$-field, for the following reason.
According to \ewitten, on  a spacetime $X$, the general flux quantization law
for $G$ is not that $G/2\pi$ has integral periods but
that the periods of $G/2\pi$ coincide with the periods of $c_2(X)/2$
mod integers.  (There is a slightly more general formulation if $X$
is not a complex manifold.)
In our situation, the integral of $c_2(X)/2$ over ${\bf CP}^2$
is a half-integer,\foot{Let the total Chern class of the tangent
bundle of ${\bf CP}^2$
be $1+c_1+c_2$.  The total Chern class of the cotangent bundle
of ${\bf CP}^2$ is then $1-c_1+c_2$.  The total Chern class
of $T^*{\bf CP}^2$, restricted to ${\bf CP}^2\subset T^*{\bf CP}^2$,
is hence $(1-c_1+c_2)(1+c_1+c_2)=1-c_1^2+2c_2$, so
$c_2(T^*{\bf CP}^2)=-c_1^2+2c_2$.  Since $\int_{{\bf CP}^2}c_1^2=9$,
which is odd, the claim follows.}
so we need
\eqn\ucnon{\int_{{\bf CP}^2}{G\over 2\pi}\in \Z+{1\over 2},}
and in particular $G$ cannot be zero.

If we normalize the four-form $\alpha$ to represent the class
$[{\bf CP}^2]$, then $\alpha$ generates $H^4_{cpct}(X;\Z)$ (or rather
its image in real cohomology).
Also, $\alpha\cdot \alpha = 3$, so the dual lattice $H^4(X;\Z)$
is generated by $\alpha/3$.  Hence, we require
\eqn\buchon{\left[{G\over 2\pi}\right]
={\alpha\over 3}\left(k+\half\right)~~~{\rm with}~
k\in \Z.}
One also has $H^4(Q;\Z)=H^4(X;\Z)/H^4_{cpct}(X;\Z)=\Z_3$.
The different possibilities for the restriction of the $C$-field to
$\partial X=Q$ are determined by the value of $k$ modulo three.

In the presence of $N$ membranes and a $G$-field, the membrane flux
at infinity is
\eqn\xoxn{\Phi=N+{1\over 2}\int_X{G\wedge G\over (2\pi)^2}
= N+{(k+\half)^2\over 6}.}
In evaluating the integral, we used \buchon\ and the fact that
$\alpha\cdot \alpha=3$.  A check on \xoxn\ is that if $k$ is shifted
by an integer multiple of 3 (the 3 is needed so as to leave fixed the
restriction of $G$ to $Q$), $\Phi$ changes by an integer.
According to the discussion in section 2, a model is specified
by fixing the value of $\Phi$ and also by fixing the value of $k$ modulo
3.  A supersymmetric vacuum is then found by finding a nonnegative
$N$ and an integer $k$ in the given mod 3 coset such that \xoxn\ is obeyed.
There is precisely one case of a model having more than one vacuum,
with all vacua having $N=0$.  This arises for $\Phi=3/8$,
with $k=1$ and $k=-2$.  We do not know a Landau-Ginzburg or
other semiclassical description for this ${\cal N}=3$ model
with two vacua (but see below).

For sufficiently large $\Phi$, this model (at $\vec d=0$)
is expected to flow to a nontrivial
superconformal field theory in the infrared.  Indeed, the standard
conjectures would suggest that the SCFT in question is dual to
$M$-theory on $AdS_3\times Q$, with the $C$-field on $Q$ being
determined by the value of $k$ modulo three.  We have no good way
at present to determine if the model flows to a nontrivial SCFT
also for small $\Phi$.

We expect that instead of $T^*{\bf CP}^2$, one could in a similar
way analyze $T^*F$, where $F$ is a two-dimensional Fano surface.
One can also consider a collection of intersecting ${\bf CP}^2$'s
(with a suitable normal bundle over it) and carry out a similar analysis.

\bigskip\noindent{\it Physical Interpretation Of Gauge Theory?}

So far the $U(1)$ gauge theory with three charged hypermultiplets
has been considered just as a mathematical device. It is natural
to wonder whether, in fact, this gauge theory can be interpreted
physically as the long wavelength theory of a membrane probe
of the $\R^3\times T^*{\bf CP}^2$ solution of $M$-theory.
More generally, we would like to find an effective action for $N$ membranes
 probing the $\R^3\times T^*{\bf CP}^2$ singularity (in the
limit that ${\bf CP}^2$ is ``blown down'') that will give
a gauge theory dual of $M$-theory on ${\bf AdS}_4\times Q$.
In the spirit of \refs{\kleb,\morrison}, such a description
might be roughly as follows.  Consider an ${\cal N}=4$ supersymmetric
 gauge theory in three dimensions with gauge group $S(U(N)\times U(N))$
and hypermultiplets consisting of three copies of $({\bf N},\bf {\bar N})$.
Break ${\cal N}=4$ to ${\cal N}=3$ with some Chern-Simons interaction,
determined by the $C$-field.
 (Gauge theories with Chern-Simons interactions are
essentially the only known classical field theories in three spacetime
dimensions without gravity
with ${\cal N}=3$ supersymmetry.  For a study of their
dynamics in the abelian case, see \ref\strkap{A. Kapustin and M.
Strassler, ``On Mirror Symmetry In Three-Dimensional Abelian Gauge
Theories,'' JHEP {\bf 9904:021,1999}, hep-th/9902033.}.)  Such
a model might have roughly the right properties.

\subsec{Blowup Of Orbifold Singularity}

The other kind of singularity that we will briefly examine is
a simple orbifold singularity.  We begin with $\C^4$, with
complex coordinates $z_1,\dots,z_4$,
and consider the $\Z_4$ symmetry $z_a\to iz_a$.  The quotient
$\C^4/\Z_4$ is a Calabi-Yau orbifold.

If one analyzes this type of orbifold in Type IIA string theory,
one finds that there is one blow-up mode and no complex structure
deformation.  The blow-up corresponds to a very simple resolution
of the singularity, in which it is replaced by the total space
$W$ of a line bundle ${\cal L}={\cal O}(-4)$ over ${\bf CP}^3$.
Thus, ${\bf CP}^3$ is embedded in $W$ as an exceptional divisor,
the ``zero section'' of ${\cal L}$.  $W$ admits a Calabi-Yau metric,
asymptotic in closed form to the flat metric on $\C^4/\Z_4$;
because of the $SU(4)$ symmetry of $W$, it is actually possible
to describe this metric by quadrature, though we will not do so here.

One might at first think that one could approach the $\C^4/\Z_4$
orbifold singularity in $M$-theory by a motion in K\"ahler moduli space,
leading to a blow-down of the exceptional
divisor ${\bf CP}^3\subset W$.
However, since the Hodge numbers $h^{i,0}({\bf CP}^3)$ are zero
for $i>0$, fivebrane wrapping on ${\bf CP}^3$ will produce
a superpotential \ugwitten, proportional roughly to $e^{-V}$
with $ V$ the volume of ${\bf CP}^3$.  Moreover, though the multiple cover
formula for multiple fivebrane wrapping in $M$-theory is not known,
analogy with other multiple cover formulas (such as the formula
\nref\cand{P. Candelas, X.C. De la Ossa, P.S. Green and L. Parkes,
``A Pair of Calabi-Yau Manifolds As An
Exactly Soluble Superconformal Theory,'' Nucl. Phys. {\bf B359}
(1991) 21.}%
\nref\aspmor{P. Aspinwall and D. Morrison,
``Topological Field Theory and Rational Curves,''
Commun. Math. Phys. {\bf 151} (1993) 245.}%
for multiple covers by fundamental strings \refs{\cand,\aspmor})
suggests that the sum over multiple covers of ${\bf CP}^3$ will
produce a pole at $V=0$.  If this is so, there will not be interesting
long distance physics associated with the behavior of $M$-theory near
a $\C^4/\Z_4$ singularity.   At any rate, one certainly cannot
expect to study $M$-theory on $\C^4/\Z_4$ while ignoring the superpotential.

Is it possible to include a $G$-field on $W$ while preserving supersymmetry?
If  so, then since the $G$-field must vanish in cohomology
on a fivebrane worldvolume (because of the existence of a field $T$
on the fivebrane with $dT=G$), in the presence of the $G$-field
the superpotential would be absent, and the question of the behavior
near the $\C^4/\Z_4$ singularity would be restored.

The answer to the question of whether a supersymmetric $G$-field
is possible turns out, however, to be ``no,'' in the following
interesting way.  First of all,  $W$ is contractible to ${\bf CP}^3$,
so its  nonzero Betti numbers
are $h^0=h^2=h^4=h^6=1$.  For cohomology with compact support,
one has the dual Betti numbers $h^2_{cpct}=h^4_{cpct}=h^6_{cpct}
=h^8_{cpct}=1$.  This suggests that the map $i:H^k_{cpct}(W;\R)
\to H^k(W;\R)$ may be nonzero for $k=2,4,$ and 6.  A topological
analysis, using the fact that $c_1({\cal L})^3\vert_W\not=0$,
shows that this is so.  Consequently, given the asymptotically
conical nature of the Calabi-Yau metric on $W$, we expect the
space of ${\bf L}^2$ harmonic forms on $W$ to be three-dimensional,
with one class each in dimension 2, 4, and     6.  Given this, there
is only one option for how the $SU(2)$ group $R$ that acts on the
cohomology of a K\"ahler manifold (see Appendix I) can act on the
${\bf L}^2$ harmonic forms on $W$: they transform with spin 1.  Hence,
though there is an ${\bf L}^2$ harmonic four-form on $W$, it is not
primitive, and one cannot turn on a $G$-field without breaking supersymmetry.

\newsec{Brane Perspective}

We will conclude this paper by pointing out a reinterpretation
of the problem in terms of singularities of branes.
We already explained in section 2.6 that $F$-theory on a four-fold
singularity can be reinterpreted as Type IIB with a D7-brane
that has a world-volume $\R^4\times L$, where $L\subset \C^3$ is
a singular complex surface.
By successive circle compactifications, it follows that
$M$-theory or Type IIA at a four-fold singularity can be
described by Type IIA with a singular sixbrane $\R^3\times L$,
or Type IIB with a singular fivebrane $\R^2\times L$.

Analogous phenomena have been noted in the past in the context of
${\cal N}=2$ conformal theories with NS or M5-branes worldvolumes with
singular geometry $\R^4\times \Sigma$ where $\Sigma$
is a Riemann surface which develops a singularity, say of the
form $y^2=x^n$ locally (for $n>2$).  These give models
for studying Type II strings at a Calabi-Yau threefold singularity capturing
Argyres-Douglas points of ${\cal N}=2$ conformal theories \ref\ad{P.C.
Argyres and M.R. Douglas,``New Phenomena in SU(3) Supersymmetric
Gauge Theory,'' Nucl. Phys. {\bf B448} (1995) 93.}\
and are presently under study \shva .

As we discussed in section 2.6, it is important that in the cases
that we have looked at, there are no corrections to the classical
$\R^n\times L$ geometry.  Let us raise the general question of this sort.
  Suppose we have a
$p$-brane of some kind, with worldvolume
$$\R^n\times X^{p+1-n}$$
embedded in $\R^{10}$  or $\R^{11}$ depending on whether
we are dealing with string theory or $M$-theory. We assume
that this geometry preserves some number of supersymmetries in $\R^n$.
Let us assume $X$ develops a singularity.
Is this singular geometry  smoothed
out in the quantum theory?  A necessary condition for that is
 the existence of instantons which end on $X$.  So if $q$-branes
 can end on this particular $p$-brane, the condition
 is the absence of compact $q$-cycles in $X$.  So for
 $Dp$-branes in  Type IIA or IIB string theory,
 since $D(p-2)$-branes and fundamental one-branes
 can end on them, the condition is the absence of topologically
 non-trivial compact
 one-cycles and $p-2$-cycles in the geometry of $X$.  For $M5$-branes,
  the condition is the absence of compact two-cycles in
 $X$.

\appendix{I}{Primitive Forms}

The de Rham cohomology of a compact K\"ahler manifold $X$,
or the space of ${\bf L}^2$ harmonic forms on any K\"ahler manifold,
admits an $SU(2)$ action which is as follows.  (See \ref\griff{P. Griffiths
and J. Harris, {\it Principles Of Algebraic Geometry} (Wiley-Interscience,
1978).}, pp. 122-6, for a mathematical introduction.)  A diagonal
generator $J_3$ of $SU(2)$ multiplies a $p$-form by $(n-p)/2$,
where $n$ is the complex dimension of $X$.  The lowering operator
$J_-$ acts by wedge product with the K\"ahler form $K$:
\eqn\umby{G\to K\wedge G.}
And the raising operator $J_+$
is the adjoint operation of contraction with $K$:
\eqn\jumby{G_{i_1i_2\dots i_n}\to K^{i_1i_2}G_{i_1i_2\dots i_n}.}

Conceptually, this $SU(2)$ action arises as follows.  Begin
with the supersymmetric nonlinear sigma model in four dimensions
with target space $X$, and dimensionally reduce it to $0+1$ dimensions.
This gives a supersymmetric system in which the Hilbert space is the
space of differential forms on $X$, the four supercharges are
are the $\partial$ and $\bar\partial$ operators and their adjoints,
and there is an $SU(2)$ symmetry that comes from rotations of the
three extra dimensions.   From this point of view, the $SU(2)$ arises
as an $R$-symmetry group, so we denote it as $R$.

Since an $(n-p)$-form has $J_3$ eigenvalue  $(n-p)/2$, it clearly
transforms under $R$ with spin at least $|n-p|/2$.
For $n-p\geq 0$, we declare the primitive part of $H^{n-p}(X;{\bf R})$
to consist of the harmonic forms that transform with spin precisely
$(n-p)/2$.  For a middle-dimensional form, with $p=n$, this
definition means that the primitive part of $H^n(X;\R)$ consists
precisely of the $R$-invariants.

For a noncompact K\"ahler manifold $X$, if all of the ${\bf L}^2$ harmonic
forms are in the middle dimension, then they are all automatically
primitive.  For a middle-dimensional ${\bf L}^2$ harmonic form
that is not $R$-invariant can be raised and lowered to make
${\bf L}^2$ harmonic forms of other dimensions.

For a middle-dimensional ${\bf L}^2$ harmonic form $G$, such
as the $G$-field on a Calabi-Yau four-fold, primitiveness
is equivalent to either $0=J_-G$, which is the condition on $G$
given in \beckers, or $0=J_+G=K\wedge G$.


An important illustrative case is that of a complex surface $W$.
The middle-dimensional cohomology of $W$ is two-dimensional and
can be decomposed as follows.  The space of self-dual forms at a given
point is three-dimensional; the self-dual forms are the $(2,0)$ and
$(0,2)$ forms and the multiples of the K\"ahler class $K$.
The $(2,0)$ and $(0,2)$ forms are clearly primitive (the lowering
operator would map them to $(3,1)$ and $(1,3)$-forms) but the K\"ahler
class $K$ is not (as $K\wedge K\not= 0 $).  The anti-self dual two-forms
are of type $(1,1)$ and are of
the form $\alpha = a_{i\bar j} dz^i\wedge d\bar z^{\bar j}$ where
$a_{i\bar j}$ is traceless.  Tracelessness of $a$ means that $\alpha$
is annihilated by the raising and lowering operators and so is primitive.
  So for a complex surface,
the middle-dimensional
primitive cohomology is of type $(2,0)$ or $(0,2)$ and self-dual,
or of type $(1,1)$ and anti-self-dual.

Closer to our needs in this paper is the case of
 a complex four-fold $X$.
  At any point $P\in X$, the holonomy group $U(4)$ acts
on the differential forms at $P$.  Since the generator of the center
of $U(4)$ simply multiplies a $(p,q)$ form by $p-q$, we focus on the
$SU(4)$ action.  We look first at the $(p,p)$ forms for $p=0,1,2,\dots$,
since they are closed under the action of $R$.
The $(0,0)$ forms transform in the trivial representation ${\bf 1}$ of
$SU(4)$.  The $(1,1)$-forms $a_{i\bar j}dz^i\wedge dz^{\bar j}$ transform
as ${\bf 4}\otimes \overline{\bf 4}={\bf 1}\oplus {\bf 15}$, with
${\bf 15}$ the adjoint representation.  Since a $(2,0)$ or $(0,2)$-form
$h_{ij}dz^i\wedge dz^j$ or $\tilde h_{\bar i\bar j}d\bar z^{\bar i}
\wedge d\bar z^{\bar j}$ transforms as the ${\bf 6}$, the $(2,2)$-forms
transform as ${\bf 6}\otimes {\bf 6}={\bf 1}\oplus {\bf 15}\oplus {\bf 20}$.
{}From this, it follows that $(2,2)$-forms that transform as ${\bf 20}$ of
$SU(4)$ have $R=0$, those that transform as ${\bf 15}$ have $R=1$,
and those that transform as ${\bf 1}$ have $R=2$.

In particular, the primitive $(2,2)$-forms transform in an irreducible
representation of $SU(4)$.  From this, it follows that they all
transform with the same eigenvalue under the Hodge $*$ operator.
To determine the sign, it suffices to consider the case that $X=W_1\times W_2$,
with the $W_i$ complex surfaces, and to consider on $X$ the primitive
$(2,2)$-form $G=\alpha_1\wedge \alpha_2$, where for $i=1,2$,
$\alpha_i$ is a primitive $(1,1)$-form on $W_i$.  Since the $\alpha_i$
are anti-self-dual, $G$ is self-dual.

We can similarly analyze the primitive $(3,1)$ cohomology.  The $(2,0)$-forms
transform as ${\bf 6}$ under $SU(4)$, while the $(3,1)$-forms transform
as ${\bf 6}\oplus {\bf 10}$.   $(3,1)$-forms that transform as ${\bf 10}$
of $SU(4)$ have $R=0$ and so are primitive, while those that transform
as ${\bf 6}$ have $R=1$.  Since the primitive
$(3,1)$-forms transform irreducibly
under $SU(4)$, they again all have the same eigenvalue of $*$.
Indeed, by considering the case $G=\alpha\wedge \beta$, with $\alpha$
a
primitive $(1,1)$-form on a surface $W_1$ and $\beta$ a primitive
$(2,0)$-form on another surface $W_2$, we learn that the primitive
$(3,1)$ cohomology of a four-fold is anti-self-dual.

Finally, the $(4,0)$ cohomology of a four-fold transforms
trivially under $SU(4)$ and is primitive.  By setting $G=\beta_1\wedge \beta_2$
with $\beta_i$ a $(2,0)$ form on $W_i$ for $i=1,2$, we learn that
the $(4,0)$ cohomology on a four-fold is self-dual.

In sum, the Hodge $*$ operator acts on the primitive $(p,4-p)$ cohomology
of a four-fold as $(-1)^p$.   This type of argument can clearly
be generalized to other dimensions.

\appendix{II}{Supersymmetry Conditions in Eleven Dimensions}

In this appendix we extend the analysis of \beckers\ to compactifications
of M-theory on Calabi-Yau four-folds with general $G$-flux,
allowing for the possibility
that the three-dimensional cosmological constant is non-zero.
In the original version of this paper we showed that turning on
holomorphic $G$-flux of type $(4,0)$ induces a mass
for the gravitino fields, $m_{\psi} \sim \int G \wedge \Omega$.
In a supersymmetric situation masses of the bosonic modes of
the supergravity multiplet should be related to $m_{\psi}$ in
a supersymmetric fashion. In particular, one should expect a non-zero
cosmological constant $\Lambda = - \vert m_{\psi} \vert^2$.
However, it turns out that compactification with a $G_{4,0}$
flux leads to a solution with zero cosmological constant and,
therefore, implies broken supersymmetry, see below
and \ref\Polchinski{S.~Giddings, S.~Kachru, and J.~Polchinski, to appear.}.

We follow the notations of \beckers\ where capital letters
$M$, $N$, $\ldots$ run from 0 to 10 and denote eleven-dimensional
indices; $m$, $n$, $\ldots$ are real indices tangent to $Y$;
and Greek letters $\mu$, $\nu$, $\ldots$ stand for the
three-dimensional Lorentzian indices 0,1,2.
Finally, lower case letters $a$, $b$, $\ldots$ from the beginning
of the alphabet denote holomorphic indices tangent to $Y$.

The bosonic part of the eleven-dimensional effective action,
corrected by the $\sigma$-model anomaly on the five-brane
world-volume, has the following form:
\eqn\maction{ S_{11} = {1 \over 2} \int d^{11}x \sqrt{-g} R -
{1 \over 2} \int \Big[ {1 \over 2} G \wedge * G +
{1 \over 6} C \wedge G \wedge G - (2 \pi)^4 C \wedge I_8 \Big] }
The eight-form anomaly polynomial can be expressed in
terms of the Riemann tensor \ref\AGW{L.~Alvarez-Gaum\'e and E.~Witten,
Nucl.Phys. {\bf B234} (1983) 269.}:
\eqn\anomaly{
I_8 = {1 \over (2 \pi)^4} \Big( - {1 \over 768} ({\rm tr} R^2)^2
+ {1 \over 192} {\rm tr} R^4 \Big)}
In these units the five-brane tension $T_6 = {1 \over (2 \pi)^3 }$.
The field equation for $G$
that follows from the action \maction\ looks like:
\eqn\eom{d * G = - {1 \over 2} G \wedge G + (2 \pi)^4 I_8
- \sum_{i=1}^N \delta^8 (x^m - P_i)}
This equation is a macroscopic analog of the anomaly equation \sethrel.
In fact, the right-hand side of \eom\ represents a local source for the field
$G_{\mu \nu \rho m}$. For a compact Calabi-Yau space $Y$, the $G$-flux
has nowhere to go. Hence, the integral of the right-hand side of \eom\
has to vanish, leading to the anomaly cancellation condition \sethrel.
In order to satisfy the equation of motion \eom\
we take the following ansatz for the three-dimensional components of $G$:
\eqn\gfield{G_{\mu \nu \rho m} =
\epsilon_{\mu \nu \rho} \partial_m f(x^m)}

We also allow for arbitrary internal components $G_{mnpq}$,
the form of which will be fixed by the field equation and
supersymmetry conditions.  As we will see in a moment,
a maximally symmetric compactification on $Y$
with nontrivial $G$-flux typically leads to warped metric:
\eqn\mmetric{ds_{11}^2 = \Delta^{-1} \Big( ds_3^2(x^{\mu}) + ds_8^2 (x^m)
\Big) }
where we introduced the warp factor $\Delta (x^m)$.
For now, both $\Delta(x^m)$ and $f(x^m)$ are scalar
functions of the coordinates on $Y$. Below we show that
these two functions are related by the supersymmetry conditions
which we are going to analyze now.

Assuming that the gravitino, $\psi_M$,
vanishes in the background, supersymmetry variations of
the bosonic fields are all identically zero. So,  we only have
to check that the variations of the gravitino also vanish
for some Majorana spinor $\eta$:
\eqn\msusy{ \delta \psi_M \equiv
\nabla_M \eta - {1 \over 4} {\Gamma_M}^N \partial_N (\log \Delta) \eta -}
$$
- {1 \over 288} \Delta^{3/2}
({\Gamma_M}^{PQRS} - 8 \delta_M^P \Gamma^{QRS}) G_{PQRS} \eta =0
$$
The first two terms in this expression come from the covariant
derivative in the eleven-dimensional metric \mmetric.

Following \beckers, we make the 11=3+8 split:
$$
\Gamma_{\mu} = \gamma_{\mu} \otimes \gamma_9,
\quad \Gamma_m = 1 \otimes \gamma_m
$$
where the eleven-dimensional
gamma-matrices $\Gamma^M$ are hermitian for $M=1,\ldots, 10$
and anti-hermitian for $M=0$. They satisfy:
\eqn\anticomm{ \{ \Gamma_M , \Gamma_N \} = 2 g_{MN}}
We use the standard notation
\eqn\asymgamma{\Gamma_{M_1 \ldots M_n} = \Gamma_{[M_1}
\ldots \Gamma_{M_n]}}
for the antisymmetrized product of gamma-matrices.
We decompose the supersymmetry parameter as:
\eqn\etadecomp{ \eta = \epsilon \otimes \xi + \epsilon^* \otimes \xi^*}
where $\epsilon$ is an anti-commuting Killing spinor in three dimensions:
\eqn\eps{ \nabla_{\mu} \epsilon = m_{\psi} \gamma_{\mu} \epsilon^*}
%
and $\xi$ is a commuting eight-dimensional complex spinor
of definite chirality. From the commutation relation of
$\nabla_{\mu}$ in anti-de Sitter space we find the usual
relation between the complex gravitino mass $m_{\psi}$
and the value of the three-dimensional cosmological constant:
\eqn\cosmc{\Lambda = - \vert m_{\psi} \vert^2}

Without loss of generality we can take:
\eqn\xichiral{\gamma_9 \xi = \xi}
Here $\gamma_9$ is the eight-dimensional chirality operator that
anti-commutes with all the $\gamma_m$'s and satisfies $\gamma_9^2=1$.
The sign in \xichiral\ determines whether it is space-filling
membranes or space-filling antimembranes that can be included without
breaking supersymmetry.
If the sign is changed,
 the corresponding supersymmetric vacuum can be obtained from that
with $\gamma_9 \xi = + \xi$ by changing the sign of the
function $f$ and the chirality of $\epsilon$.

The $\mu$-component of the supersymmetry condition \msusy\
takes the form:
\eqn\mmsusy{ \eqalign{\delta \psi_{\mu} &\equiv \nabla_{\mu} \eta -
{1 \over 4} \partial_n (\log \Delta)
(\gamma_{\mu} \otimes \gamma_9 \gamma^n) \eta - \cr
&
 - {1 \over 288} \Delta^{3/2}
(\gamma_{\mu} \otimes \gamma_9 \gamma^{mnpq}) G_{mnpq} \eta
+ {1 \over 6} \Delta^{3/2} (\partial_m f)
(\gamma_{\mu} \otimes \gamma^{m}) \eta =0 \cr}}
Substituting the decomposition \etadecomp\ for $\eta$
and using equations \eps\ and \xichiral\ we obtain our
first supersymmetry condition. Projecting the result
onto subspaces of positive and negative chirality, we
actually get two conditions: one comes from the first
and the third term in \mmsusy; and the other one comes from
the second and the fourth term.
Moreover, in a vacuum with ${\cal N}=2$ supersymmetry
variations proportional to $\epsilon$ and $\epsilon^*$
must vanish separately. Therefore, we find:
\eqn\msol{f= \Delta^{-3/2}, \quad \quad
288 m_{\psi} \Delta^{-3/2} \xi = {\slash\!\!\!\! G} \xi^*.}
Here we have written ${\slash\!\!\!\! G}$
for the total contraction $G_{mnpq} \gamma^{mnpq}$.
According to our ansatz \gfield, the three-dimensional
components of the four-form field strength:
\eqn\gfieldsol{ G_{\mu \nu \rho m} =
\epsilon_{\mu \nu \rho} \partial_m \Delta^{-3/2}}
have the form similar to the membrane solution with
local ``effective'' membrane charge density,
as follows from the field equation \eom\ for the internal components.
Note, that for a compact $Y$ there is no global membrane charge.
Substituting \gfieldsol\ into \eom, we obtain two additional equations.
One equation uniquely determines $G$ given its cohomology class:
\eqn\inteom{\partial_m (\Delta^{-3/2} G^{mnpq}) = {1 \over 4!}
(\partial_m \Delta^{-3/2}) \epsilon^{mnpqrstu} G_{rstu}}
It is identically obeyed if $G$ is self-dual (as we will find) and closed.
The second equation which follows from \eom\ determines
the warp factor:
\eqn\eowarp{d * d \log \Delta^{3/2} = {1 \over 2} G \wedge G -  (2 \pi)^4 I_8
+ \sum_{i=1}^N \delta^8 (x^m - P_i) }

Now we return to the original supersymmetry condition \msusy\
and consider its $m$-component:
\eqn\mmmsusy{\delta \psi_m \equiv \epsilon \otimes \nabla_m \xi
- m_{\psi} \epsilon \otimes \gamma_m \xi^* +
{1 \over 24} \Delta^{3/2} G_{mnpq} \epsilon \otimes \gamma^{npq} \xi +}
$$
+ {1 \over 4} \partial_m (\log \Delta) \epsilon \otimes \xi
- {3 \over 8} \partial_n (\log \Delta) \epsilon \otimes {\gamma_m}^n \xi
+ {\rm c.c.} =0
$$
where we used the explicit form of the solution \msol\
and standard properties of gamma-matrices.
Once again $\epsilon$ and $\epsilon^*$ components
of this equation must vanish separately.
By means of the rescaling transformations:
$$
\matrix{ g_{mn} & \to & \Delta^{3/2} g_{mn} \cr
\xi & \to & \Delta^{-1/4} \xi }
$$
the $\epsilon$-component of \mmmsusy\ can
be written in the following compact form:
\eqn\mfsusy{ \nabla_m \xi - m_{\psi} \Delta^{3/4} \gamma_m \xi^*
+ {1 \over 24} \Delta^{-3/4} G_{mnpq} \gamma^{npq} \xi = 0}

Then, following \beckers, we choose $\xi$
to be a covariantly constant spinor of unit norm, and use it to define
the complex structure ${J_m}^n = i \xi^{\dagger} {\gamma_m}^n \xi$
and the K\"ahler form $J_{a \bar b} = i g_{a \bar b}$.
Since the metric on $Y$ is of type (1,1), it is convenient to think
of `holomorphic' gamma-matrices $\gamma^a$ and $\gamma_a$
as creation and annihilation operators that satisfy the algebra:
$$
\{ \gamma^{a}, \gamma^{b} \} = \{ \gamma^{\bar a}, \gamma^{\bar b} \} = 0,
\quad \{ \gamma^{a}, \gamma^{\bar b} \} = 2 g^{a \bar b}
$$
Namely, $\gamma^a$ and $\gamma_{\bar a}$ act on
the Fock ``vacuum'' $\xi$ as annihilation operators:
\eqn\annihil{ \gamma^a \xi =0, \quad \gamma_{a} \xi^* =0,
\quad \gamma^{\bar a} \xi^* =0, \quad \gamma_{\bar a} \xi =0}

To obtain the algebraic constraints on the field $G$, we multiply
the differential equation \mfsusy\ by $\gamma^a$ which kills
the first term in that equation:
\eqn\mcond{ 24 m_{\psi} \Delta^{3/2} \gamma^a \gamma_m \xi^* -
G_{mnpq} \gamma^a \gamma^{npq} \xi = 0}
Components of this equation with different gamma-matrix structure
must vanish separately. For example, if we choose $m$ to be
an anti-holomorphic index and use \annihil, we find
that $G^{(1,3)}$, the (1,3) piece of the field $G$, must be zero:
\eqn\gtrione{G_{\bar a \bar b \bar c d} = 0.}
Moreover, $G^{(2,2)}$ must be primitive:
\eqn\gprim{G_{\bar a b \bar c d} J^{\bar c d} = 0}

Finally, taking the trace over the holomorphic index `$a$'
in the main supersymmetry condition \mcond, we can
demonstrate that the (4,0) part of the $G$-flux breaks
supersymmetry, as we expected in section 2.4.
Indeed, we get a relation between $G_{4,0}$ and $m_{\psi}$
of the form \msol, but with a different numerical coefficient:
$$
96 m_{\psi} \Delta^{3/2} \xi^* = G_{abcd} \gamma^{abcd} \xi
$$

Therefore, compactifications of M-theory on Calabi-Yau four-folds
with $G_{4,0} \neq 0$ lead to three-dimensional vacua with broken
supersymmetry, in accordance with the proposed expressions
\xonn\ and \kxob\ for the effective superpotential. Indeed, from \xonn\
it follows that (4,0) part of $G$ contributes to the vacuum value of $W$.
Now, in order to see that three-dimensional vacua with $W \neq 0$
are not supersymmetric we must be careful and use the appropriate
covariant derivatives $D/D t_i$ and $D/D k_j$.
Since the supersymmetry conditions \juxin\ and \lully\ are scale invariant,
there is at least one massless vector multiplet in the effective
three-dimensional theory corresponding to the volume of $Y$.
After this multiplet is dualized to a massless chiral
multiplet $K_0$, one finds no-scale supergravity theory
where effective superpotential is independent on $K_0$,
see \Polchinski, for otherwise this multiplet had a mass.
Since $\partial_{k_0} W = 0$ but $\partial_{k_0} K(t_i, k_j) \neq 0$,
where $K(t_i, k_j)$ is the K\"ahler potential, it follows that
in a vacuum with $W \neq 0$ the covariant derivative
$DW/D k_0 = \partial_{k_0} W + (\partial_{k_0} K) W$
does not vanish, and supersymmetry is broken.

Note, that both $G_{4,0}$ and primitive $G$-flux of type $(2,2)$
are self-dual and, therefore, obey equations of motion \inteom.
As one can easily see, self-duality of $G$ also implies vanishing
of the three-dimensional cosmological constant.
Indeed, it follows from \sethrel\ that
${1 \over 2} \int G \wedge ^* G - \chi (X)/24 =0$,
so that the positive vacuum energy due to the $G$-flux
is cancelled by the negative contribution of
the $R^4$ terms in the effective action of M theory.
In particular, it means that compactification with $G$-flux
of type $(4,0)$ classically gives a flat space solution with $\Lambda =0$,
{\it cf.} \ref\HLouis{M.~Haack, J.~Louis, ``M-theory compactified
on Calabi-Yau fourfolds with background flux," hep-th/0103068.}.

\bigskip\noindent
{\bf Acknowledgements} We have benefited from discussions
with M. Bershadsky, N. Hitchin, S. Katz, J. Louis, R. MacPherson,
D. Morrison, T. Pantev, N. Seiberg, A. Shapere, and S.-T. Yau.
We thank J.~Polchinski hor helpful and interesting discussions
that stimulated a revision of the analysis of compactifications
with $G_{4,0} \neq 0$ in Appendix II.

The work of S.G. was supported in part by grant RFBR
No 98-02-16575 and Russian President's grant No 96-15-96939.
The work of C.V. was supported in part by NSF grant PHY-98-02709
and that of E.W. was supported in part by NSF grant PHY-9513835.

\listrefs
\end